\documentclass{article}
\usepackage{graphicx}  
\usepackage{amsmath}   
\usepackage[compress]{cite}
\usepackage{amssymb}   
\usepackage{bm} 
\usepackage{dcolumn}
\usepackage{color}
\usepackage{mathrsfs}
\usepackage{amsfonts}
\usepackage[caption=false]{subfig}
\usepackage{varioref}
\usepackage{textcomp}
\usepackage{tikz}
\usepackage{tikz,pgfplots}
\usepackage[utf8]{inputenc}
\usepackage{multicol}
\usepackage[english]{babel}
\usepackage{amsmath}
\usepackage{multirow}
\usepackage{graphics}
\usepackage{colortbl}
\usepackage{colordvi}
\usepackage{verbatim}
\usepackage{mathtools}
\usepackage{amsmath}
\usepackage{braket}
\usepackage{tikz}
\usepackage{amssymb}
\usepackage{slashed}
\usepackage{color,soul}
\setul{0.5ex}{0.3ex}
\setulcolor{red}
\usepackage{diagbox}
\usepackage{newcent}

\RequirePackage[colorlinks,citecolor=blue,urlcolor=magenta,linkcolor=blue]{hyperref}
\allowdisplaybreaks
\def\KR{Kalb-Ramond }
\def\a {\alpha}
\def\b{\beta}
\def\m{\mu}
\def\n{\nu}

\def\gr{general relativity}

\labelformat{section}{Section #1} 
\labelformat{subsection}{Section #1} 
\labelformat{subsubsection}{Section #1}
\labelformat{subsubsubsection}{Section #1}
\labelformat{equation}{Eq.~(#1)} 
\labelformat{figure}{Fig.~#1} 
\labelformat{subfigure}{Fig.~\thefigure#1} 
\labelformat{table}{Tab.~#1} 
\labelformat{appendix}{Appendix #1}
\definecolor{ao(english)}{rgb}{0.0, 0.5, 0.0}
\definecolor{brightgreen}{rgb}{0.4, 1.0, 0.0}
\definecolor{darkpastelgreen}{rgb}{0.01, 0.75, 0.24}
\definecolor{darkpastelblue}{rgb}{0.47, 0.62, 0.8}
\definecolor{darkpastelpurple}{rgb}{0.59, 0.44, 0.84}
\definecolor{flame}{rgb}{0.89, 0.35, 0.13}
\title{Imprints of the Janis-Newman-Winicour spacetime on observations related to shadow and accretion}
\author{\hspace{-2.4cm} Subhadip Sau\footnote{tpss2@iacs.res.in}~,
Indrani Banerjee\footnote{tpib@iacs.res.in}~
and
and Soumitra SenGupta\footnote{tpssg@iacs.res.in}\\
{\small{\hspace{-1.5cm}School of Physical Sciences, Indian Association for the Cultivation of Science,
2A \& 2B Raja S. C. Mullick Road, Kolkata-700032, India}}}

\date{ } 
\begin{document}
\maketitle 
%%%%%%%%%%%%%%%%%%%%%%%%%%%%%%%%%%%%%%%%%%%%%%%%%%%%%%%%%%%%%%%%%%%
\begin{abstract}
The final fate of gravitational collapse of massive stars has been a subject of interest for a long time since such a collapse may lead to black holes and naked singularities alike. Since, the formation of naked singularities is forbidden by the cosmic censorship conjecture, exploring their observational differences from black holes may be a possible avenue to search for these exotic objects. The simplest possible naked singularity spacetime emerges from the Einstein massless scalar field theory with the advantage that it smoothly translates to the Schwarzschild solution by the variation of the scalar charge.
This background, known as the Janis-Newman-Winicour spacetime is the subject of interest in this work. 
%A further advantage of working with this spacetime is that it smoothly translates to the Schwarzschild solution by the variation of the scalar charge. 
We explore electromagnetic observations around this metric which involves investigating the characteristics of black hole accretion and shadow. We compute the shadow radius in this spacetime and compare it with the image of M87*, recently released by the Event Horizon Telescope Collaboration. Similarly, we derive the expression for the luminosity from the accretion disk and compare it with the observed optical luminosity of eleven Palomar Green quasars. Our analysis indicates that the shadow of M87* and the quasar optical data consistently favor the Schwarzschild background over the Janis-Newman-Winicour spacetime. The implications of this result are discussed.

\end{abstract}
%%%%%%%%%%%%%%%%%%%%%%%%%%%%%%%%%%%%%%%%%%%%%%%%%%%%%%%%%%%%%%%%%%%
\section{Introduction}
\label{Intro}
One of the classic unresolved problems in general relativity is the ultimate fate of the gravitational collapse of a massive body, such as a star. It has been conjectured that the end state of any generic complete gravitational collapse leads to a Kerr black hole characterized by only its mass and angular momentum.
All other information regarding the initial conditions of the collapse, the symmetries and the nature of matter fields that were present in the beginning of the collapse gets radiated away. 
It turns out that it is very difficult to prove this conjecture either analytically or numerically and therefore one cannot definitively say that the ultimate fate of a gravitational collapse always leads to the formation of a black hole. 
In fact, investigations reveal that such gravitational collapse with a set of allowed initial conditions often lead to the formation of naked singularities \cite{Joshi:1993zg,Waugh:1988ud,Christodoulou:1984mz,Eardley:1978tr,Ori:1987hg,Giambo:2003fd,Shapiro:1991zza,Lake:1991qrk,Goswami:2006ph,Joshi:2001xi,Harada:1998cq}, even though such objects are forbidden by the cosmic censorship conjecture \cite{Penrose:1969pc}.

While the end products of gravitational collapse continues to be debatable, it is worth exploring the observational differences between black holes and naked singularities, assuming that they have been formed by some mechanism. Given the surfeit of data available in the electromagnetic domain, this has intrigued researchers worldwide since such a study can enhance our understanding regarding the nature of compact objects at the galactic centres or in the X-ray binaries. Observations related to accretion disks \cite{Kovacs:2010xm,Blaschke:2016uyo,Stuchlik:2010zz,Bambi:2009bj,Joshi:2011zm,Pugliese:2011xn,Pradhan:2010ws} or gravitational lensing \cite{Hioki:2009na,Yang:2015hwf,Takahashi:2004xh,Gyulchev:2008ff,Virbhadra:2002ju,Virbhadra:1998dy,Virbhadra:2007kw} have revealed that black holes and naked singularities often exhibit strikingly different properties which can be used as a possible probe to differentiate between them.
Further, ultra high energy collisions and fluxes of the escaping collision products can be another possible tool to discern between the two different entities \cite{Patil:2011yb}. There are however cases when certain wormhole spacetimes and naked singularities exhibit similar observational features like that of a black hole which makes the differentiation quite difficult \cite{Shaikh:2018kfv,Cunha:2018gql,Gyulchev:2018fmd,Nedkova:2013msa}. However, this will be kept outside the purview of the present discussion.

In the present work we consider the Janis-Newman-Winicour (JNW) naked singularity which represents an exact solution of the Einstein's equations with a massless scalar field \cite{Janis:1968zz}. This solution was originally derived by Fisher \cite{Fisher:1948yn} in a different parametrization while Bronnikov \& Khodunov \cite{Bronnikov:1979uz} subsequently studied its stability. It was later rediscovered by Wyman \cite{Wyman:1981bd} and the equivalence of the Wyman solution with the Janis-Newman-Winicour spacetime was established by Virbhadra \cite{Virbhadra:1997ie}. It is interesting to note that addition of the massless scalar field in the action changes the nature of the spherically symmetric and asymptotically flat exact metric solution from the Schwarzschild black hole to the JNW naked singularity. Consequently,it can be shown that one can recover the Schwarzschild metric from the JNW spacetime by continuously adjusting a single metric parameter representative of the scalar charge of the naked singularity.

There exists several works in the literature which explored the optical properties of the Janis-Newman-Winicour spacetime, e.g. gravitational lensing and relativistic images \cite{Virbhadra:2007kw,Virbhadra:1998dy,Virbhadra:2002ju,Gyulchev:2008ff,Gyulchev:2019tvk}, accretion and shadow \cite{Chowdhury:2011aa,Gyulchev:2019tvk,Yang:2015hwf,Takahashi:2004xh}. The aim of this work is to explore the nature of shadow and the emission from the accretion disk around the Janis-Newman-Winicour spacetime and compare them with the available observations. The optical luminosity of eighty Palomar Green quasars and the recently released shadow of M87* are used as the observational sample for comparing the theoretical results.

The paper is organized as follows: In \ref{S2} we review the basic properties of the Janis-Newman-Winicour spacetime. We study the structure of the shadow cast by the JNW spacetime and compare it with the image of M87* in \ref{S4}. \ref{S3} serves as a quick overview over the `thin accretion disk' model proposed by Novikov \& Thorne which helps us to evaluate the accretion disk luminosity for a sample of eighty Palomar Green quasars. Subsequently we compare this with the observed luminosity of the quasars to distinguish the JNW spacetime from the Schwarzschild background. We end with a summary of our results and the concluding remarks in \ref{S5}. 

We use (-,+,+,+) as the metric convention and will work with geometrized units taking $G=c=1$.

%The success of general relativity in addressing the local solar system based observations e.g. perihelion precession of mercury, bending of light etc. is undeniable. Yet, general relativity suffers from various theoretical and observational shortcomings. This involves presence of singularities, lack of predictability beyond the Cauchy horizon, the necessity to invoke the exotic dark matter and dark energy in the galactic and the cosmological scales respectively. In this scenario, the possible resolutions might come from modifications in the gravity sector, or additions to the matter sector or 

%%%%%%%%%%%%%%%%%%%%%%%%%%%%%%%%%%%%%%%%%%%%%%%%%%%%%%%%%%%%%%%%%%%%%%%%%%%%%%%%%%%%%%%%%%%%%%%%%%%
%%%%%%%%%%%%%%%%%%%%%%%%%%%%%%%%%%%%%%%%%%%%%%%%%%%%%%%%%%%%%%%%%%%%%%%%%%%%%%%%%%%%%%%%%%%%%%%%%%%
\section{Janis-Newman-Winicour spacetime: A quick review }
\label{S2}
%%%%%%%%%%%%%%%%%%%%%%%%%%%%%%%%%%%%%%%%%%%%%%%%%%%%%%%%%%%%%%%%%%%

In this work we consider the Einstein massless scalar (EMS) field theory such that the massless scalar field is minimally coupled to gravity. 
The associated action is given by
\begin{equation}
S=\int d^{4}x \sqrt{-g}\left[\dfrac{R}{2\kappa^{2}}-\dfrac{1}{2}\partial_{\m}\phi(r)\partial^{\m}\phi(r)\right]
\end{equation}
where, g and R are respectively, determinant of the metric tensor and the Ricci scalar, $\kappa^{2}=8\pi G$ (G is the four-dimensional gravitational constant) and $\phi(r)$ is the minimally coupled scalar field.
In four dimension, the corresponding Einstein's gravitational field equations derived from the above action has an exact static and spherically symmetric solution\cite{PhysRevLett.20.878,PhysRevD.24.839,Virbhadra:1997ie} given by,
 \begin{flalign}\label{S2_1}
ds^{2}=-\left(1-\dfrac{b}{ r}\right)^{\gamma}dt^{2}+\left(1-\dfrac{b}{ r}\right)^{-\gamma}dr^{2}+ \left(1-\dfrac{b}{ r}\right)^{1-\gamma} r^{2}\left(d\theta^2 + \sin^{2}\theta d\phi^{2}\right)
 \end{flalign}
which is popularly known as the Janis-Newman-Winicour (JNW) solution in the literature.
In \ref{S2_1} $r$ represents the radial coordinate, $0\le \gamma\le 1$ and $b\gamma=2M$, such that the Schwarzschild metric is retrieved when $\gamma=1$.
There is a curvature singularity at $r=b$ which is also the location of the event horizon. Since the singularity is not cloaked by the event horizon this metric represents a naked singularity and hence we  confine ourselves in the region $r>b$.  
The solution for the scalar field and the associated energy-momentum tensor are respectively given by 
\begin{flalign}\label{S2_2}
\phi(r)=\dfrac{q}{b}\ln \left(1-\dfrac{b}{ r}\right) ~~~\rm{and}
 \end{flalign}
%%%%%%%%%%%%%%%%%%%%%%%%%%%%%%%
\begin{equation}\label{S2_3}
T_{\m\n}=\partial_{\m}\phi\partial_{\n}\phi-\dfrac{1}{2}g_{\m\n}\partial^{\a}\phi\partial_{\a}\phi
\end{equation}
%%%%%%%%%%%%%%%%%%%%%%%%%%%%%%%%%%%%%%%%%%%%%%%%%%%%%%%%%%%%%%%%%%%
where $b$ is related to the scalar charge $q$ by,
\begin{align}
b=2\sqrt{M^2+q^2}
\label{S2-4}
\end{align}
such that smaller $\gamma$ corresponds to a larger magnitude of the scalar field. 
%{\bf We note that in the above discussion and in what follows we have taken $G=c=1$.}   

In the context of string theory a pseudo scalar field known as the axion, arises as the dual of the field strength of the Kalb-Ramond field $B_{\m\n}$ minimally coupled to Einstein gravity in four dimensions.  
The \KR field $B_{\m\n}$ with the transformation property of a second rank anti-symmetric tensor gauge field has the following action,
\begin{align}
\label{S2_4}
S_{KR}=\int d^{4}x \sqrt{-g} \left[\dfrac{R}{2\kappa^{2}}-\dfrac{1}{12}H_{\mu\nu\alpha}H^{\mu\nu\alpha} \right]
\end{align}
where $H_{\a\m\n}=\partial_{[\a}B_{\m\n]}$ is the field strength tensor which has the pseudo-scalar axion field $H$ as its dual,
\begin{flalign}\label{S2_5}
H^{\a\m\n}=\epsilon^{\a\m\n\b}\partial_{\b}H
\end{flalign}
In terms of the axion field the energy-momentum tensor of the \KR field can be written as 
%%%%%%%%%%%%%
\begin{equation}
T_{\m\n}=\partial_{\m}H\partial_{\n}H -\dfrac{1}{2}g_{\m\n}\partial^{\sigma}H\partial_{\sigma}H
\label{S2_6}
\end{equation}
%%%%%%%%%%%%%
which resembles \ref{S2_3}.

Under a different choice of the metric ansatz, the resultant static, spherically symmetric and asymptotically flat solution of the Einstein's equations (associated with the Kalb-Ramond field) assumes a perturbative solution of the the form \cite{Kar:2002xa},

\begin{equation}
\label{9}
ds^{2}=-e^{\n(r)}dt^{2}+e^{\lambda (r)}dr^{2}+r^{2}d\Omega^{2}
\end{equation}
such that
\begin{subequations}
\begin{equation}
\label{10a}
e^{\nu(r)}=1-\dfrac{2M}{r}+\dfrac{hM}{r^{3}}+\mathcal{O}\left(\dfrac{1}{r^{4}}\right)
\end{equation}
\begin{equation}
\label{10b}
e^{-\lambda(r)}=1-\dfrac{2M}{r}+\dfrac{3h}{r^{2}}+\mathcal{O}\left(\dfrac{1}{r^{4}}\right)
\end{equation}
\end{subequations}
where $h$ refers to the axion parameter and has dimensions of $M^2$. For the solution of the Kalb-Ramond field strength and the axion field one is referred to \cite{Kar:2002xa}.
Just like the JNW space time this metric also smoothly translates to the Schwarzschild solution in the event the axion parameter $h$ vanishes.

We have already explored the properties of accretion and shadow in the spacetime with the axionic charge \cite{Banerjee:2017npv,Banerjee:2019xds}. Observational implications of several other alternative gravity models have been extensively studied in the literature \cite{Stuchlik:2008fy,Bambi:2012tg,Liu:2020ola,Liu:2020vkh,Zhu:2019ura,Chakravarti:2019aup,Chakraborty:2017qve,Visinelli:2017bny}. 
In this work, we will explore the motion of both the massless and the massive particles around the Janis-Newman-Winicour (JNW) spacetime.  
In case of massive particles we will study accretion of matter, while the properties of the black hole shadow can be investigated by studying motion of the massless particles. In both cases we will confront our theoretical findings with the available observations to provide constrain on the metric parameter $\gamma$. In each case we will compare our findings with the results obtained previously for the axion metric (\ref{9},\ref{10a},\ref{10b}).

\section{Shadow cast by the compact object governed by the Janis-Newman-Winicour spacetime}
\label{S4}
With the advent of the Event Horizon Telescope, it has been possible to obtain the image of the central compact object in the galaxy M87. This has enabled direct observations of the near
horizon regime of a black hole and has opened up a new and independent window to test
the nature of strong gravity. The shadow refers to the gravitationally lensed projection of the photon circular orbits onto the observer's sky. When light from a distant source or the surrounding accretion disk come close to the photon sphere, a part of it falls into the compact object while the remaining escapes to infinity \cite{Cunha:2018acu,Vries_1999,Gralla:2019xty,Abdujabbarov:2015xqa,Abdujabbarov:2016hnw}.  
Consequently, the observer perceives a dark patch in the local sky known as the shadow. The boundary of the shadow testifies strong gravitational lensing near the photon sphere and hence the shape and size of the shadow captures useful information regarding the nature of the background spacetime \cite{Gralla:2019xty,Bambi:2019tjh,Hioki:2009na,Vagnozzi:2019apd,Allahyari:2019jqz,Banerjee:2019nnj,Roy:2019esk}. In what follows, we will study the nature of the shadow cast by the JNW spacetime and confront it with the observed shadow of M87*. We initiate by first exploring the structure of the shadow in a most general spherically symmetric spacetime.

%%%%%%%%%%%%%%%%%%%%%%%%%%%%%%%%%%%%%%%%%%%%%%%%%%%%%%%%%%%%%%%%%%%
\subsection{Structure of the shadow in a general spherically symmetric background}

In this section, we work out the structure of the black hole shadow in a general static and spherically symmetric background given by
\begin{equation}\label{p3_General_Metric}
ds^{2}=-e^{\nu(r)}dt^{2}+e^{\lambda(r)}dr^{2}+\mathcal{R}^{2}(r) r^{2}\left(d\theta^{2}+\sin^{2}\theta d\phi^{2}\right)
\end{equation} 
This metric ansatz is a more generalized form than the one usually used in the literature due to its modified volume factor, i.e. the coefficient of $d\Omega^2$ is not just $r^2$ but also has a function of $r$ multiplied to it. This is important since we are eventually interested in studying the properties of the shadow in a metric given by \ref{S2_1}.
 
Due to the time and zenithal angle independence of the metric, the energy $E$ and the total angular momentum $L$ of 
the photons are conserved. The constants of motion are given by,  
\begin{subequations}
\begin{equation}
 E=-g_{tt}u^{t}=-p_{t}~~~\rm and
\end{equation}
\begin{equation}
L=g_{\phi\phi}u^{\phi}=p_{\phi}
\end{equation}
\end{subequations}
respectively. The Hamilton-Jacobi equation can therefore be integrated to obtain the following solution for the action,
\begin{align}
\label{13}
 S=-E t +L\phi + \bar{S}(r,\theta)
\end{align}
where $\bar{S}(r,\theta)$ is an arbitrary function of radial and angular coordinates. Assuming separability of $\bar{S}(r,\theta)$ as $\bar{S}(r,\theta)= S^{r}(r)+S^{\theta}(\theta)$, and substituting the Hamilton-Jacobi equation for $r$ and $\theta$ in the Hamiltonian we obtain,
\begin{flalign}
\mathcal{R}^2r^{2}\left( k +e^{-\n(r)}  E^2 -e^{-\lambda(r)}\left(\dfrac{dS^{r}}{dr}\right)^{2}\right)=\left(\dfrac{dS^{\theta}}{d\theta}\right)^2 +\dfrac{L^2}{\sin^2 \theta}=C+L^2 
\end{flalign}
where the separation constant $C$, known as the Carter constant represents a third constant of motion \cite{Carter:1968rr}. 
Therefore the geodesic equations for $r$ and $\theta$ are given by,
%%%%%%%%%%%%%%%%%%%%%%%%%
%%%%%%%%%%%%%%%%%%%%
\begin{flalign}
e^{\lambda +\nu}\dot{r}^{2}=-e^{\nu}\dfrac{C+L^{2}}{r^{2}\mathcal{R}^{2}}+E^{2} \equiv -V_{\rm eff}(r)+E^{2} \equiv \mathbf{R}(r)~~~\rm{and}
\end{flalign}
\begin{flalign}
\left(\mathcal{R}^{2}r^{2}\dot{\theta}\right)^{2}=C-L^{2}\cot^{2}\theta \equiv E^{2}\Theta(\theta)
\end{flalign}
%%%%%%%%%%%%%%%%%%

respectively, where
\begin{flalign}
V_{\rm eff}=e^{\nu} \dfrac{C+L^{2}}{r^{2}\mathcal{R}^{2}}
\end{flalign}
%%%%%%%%%%%%%%%%%%%%%
%\begin{align}
%\label{17}
%mathbf{R}(r)=-\frac{e^{\lambda(r)}\chi}{\mathcal{R}^{2}r^2}-\frac{e^{\lambda(r)}l^2}{\mathcal{R}^{2}r^2}+e^{\lambda(r)-\nu(r)} 
%\tag{17}
%\end{align}
%%%%%%%%%%%%%%%%%%%%%%%
represents the effective potential for radial motion of photon, while
\begin{align}\label{S3_8}
{\Theta}(\theta)=\chi-l^2 \cot^2\theta
\end{align}
such that $\chi=C/E^2$ and $l=L/E$. 
The radius of the photon sphere $r_{ph}$ is defined such that the radial velocity $\dot{r}$ vanishes and 
the effective potential $V_{\rm eff}(r)$ possesses an extrema.  Generally this turns out to be a maxima, 
representing an unstable equilibrium of the photon, resulting in either fall into the gravitating object or 
escaping to infinity due to even slight perturbation. Consequently, photon sphere plays the 
important role in determining the boundary of the shadow.

Therefore, $r_{ph}$ is obtained by solving $\mathbf{R}(r)=\mathbf{R}'(r)=V_{\rm eff}'(r)=0$, such that the above conditions yield
\begin{align}\label{S3_9}
\chi+l^2=\mathcal{R}^{2}(r_{ph})r_{ph}^2 e^{-\nu(r_{ph})}~~~~~\rm{and}
\end{align}
%%%
\begin{align}\label{S3_10}
\n'(r_{ph})=2\left[\dfrac{1}{r_{ph}}+\dfrac{\mathcal{R'}(r_{ph})}{\mathcal{R}(r_{ph})}\right]
\end{align}
respectively. The photon sphere in an arbitrary spherically symmetric metric is therefore obtained by solving \ref{S3_10}  for $r$. In the limit $\mathcal{R}=1$ we get back the known result $r\nu^\prime=2$ \cite{Banerjee:2019xds}. 

The contour of the black hole shadow in the observer's sky is obtained by considering the projection of the 
photon sphere in the image plane \cite{Bardeen:1973tla}. Determination of the shadow outline depends 
on the largest positive radius obtained by solving \ref{S3_10} \cite{Vries_1999,Cunha:2018acu}. 
Two celestial coordinates $\a$ and $\b$ which are directly related to $l$ and $\chi$ designates the 
locus of the shadow boundary \cite{Bardeen:1973tla,Vries_1999}.

%%%%%%%%%%%%%%
\iffalse
To understand this,  the metric is expressed in terms of the tetrads [$e^{\m}_{(\a)}$] for a spherically symmetric background. 
In local rest frame, the apparent velocity $v_{(\theta)}$ of the photon in the $\theta$ direction and $v_{(\phi)}$ of the photon in the $\phi$ direction , are given by,

%\begin{subequations}
\begin{align}
\label{24}
v_{_{(\theta)}}&=\dfrac{u_{\m}e^{\m}_{(\theta)}}{u_{\m}e^{\m}_{(t)}}=\dfrac{p_{\theta}e^{\theta}_{(\theta)}}{p_{t}e^{t}_{(t)}}=\dfrac{\mp \sqrt{\Theta(\theta)}e^{\n/2}}{\mathcal{R}r}~~~~~\rm{and}
\end{align}

\begin{align}
\label{25}
v_{_{(\phi)}}&=\dfrac{u_{\m}e^{\m}_{(\phi)}}{u_{\m}e^{\m}_{(t)}}=\dfrac{p_{\phi}e^{\phi}_{(\phi)}}{p_{t}e^{t}_{(t)}}=-\dfrac{le^{\n/2}}{\mathcal{R}r\sin\theta}
\end{align}
%\end{subequations}
respectively. 
For an observer in location $(r_{0},\theta_{0},0)$ the celestial coordinates are given by,
\begin{align}
\label{26}
\beta=\lim_{r_0\to\infty} r_0 v_{_{(\theta)}}(r_0,\theta_0)=\mp \sqrt{\Theta(\theta_0)} 
\end{align}
\begin{align}
\label{27}
\alpha=\lim_{r_0\to\infty} r_0 v_{_{(\phi)}}(r_0,\theta_0)=-\frac{l}{\sin\theta_0}
\end{align}
By virtue of asymptotically flat metric, expressions of celestial coordinates doesn't contain 
the information of obderver's radial position. Using \ref{S3_8} it can be shown that
\fi 
%%%%%%%%%%%%%%%%%%%%%%%%%%%5
Following the prescription as given in \cite{Vries_1999,Banerjee:2019xds}, it can be shown that
\begin{align}
\label{28}
\alpha^2 + \beta^2=\chi+l^2={r_{sh}^2}
\end{align} 
From the above analysis it can be concluded  that for any general static, spherically symmetric and 
asymptotically flat metric the shadow is circular in shape and depends on the radius of photon sphere which in turn solely depends only on the $g_{tt}$ component of the metric. We also note that for an asymptotically flat observer the radius of the shadow does not 
depend on the distance $r_0$ and the inclination angle $\theta_0$ of the observer.  

%\section{Generalised Winicour Solution and a comparison of shadow radius with perturbative solution} 

\subsection{Shadow of the compact object governed by the Janis-Newman-Winicour spacetime}
\label{S4.2}

\begin{figure}[h]
\centering
\subfloat[\label{f1a}]{\includegraphics[scale=0.68]{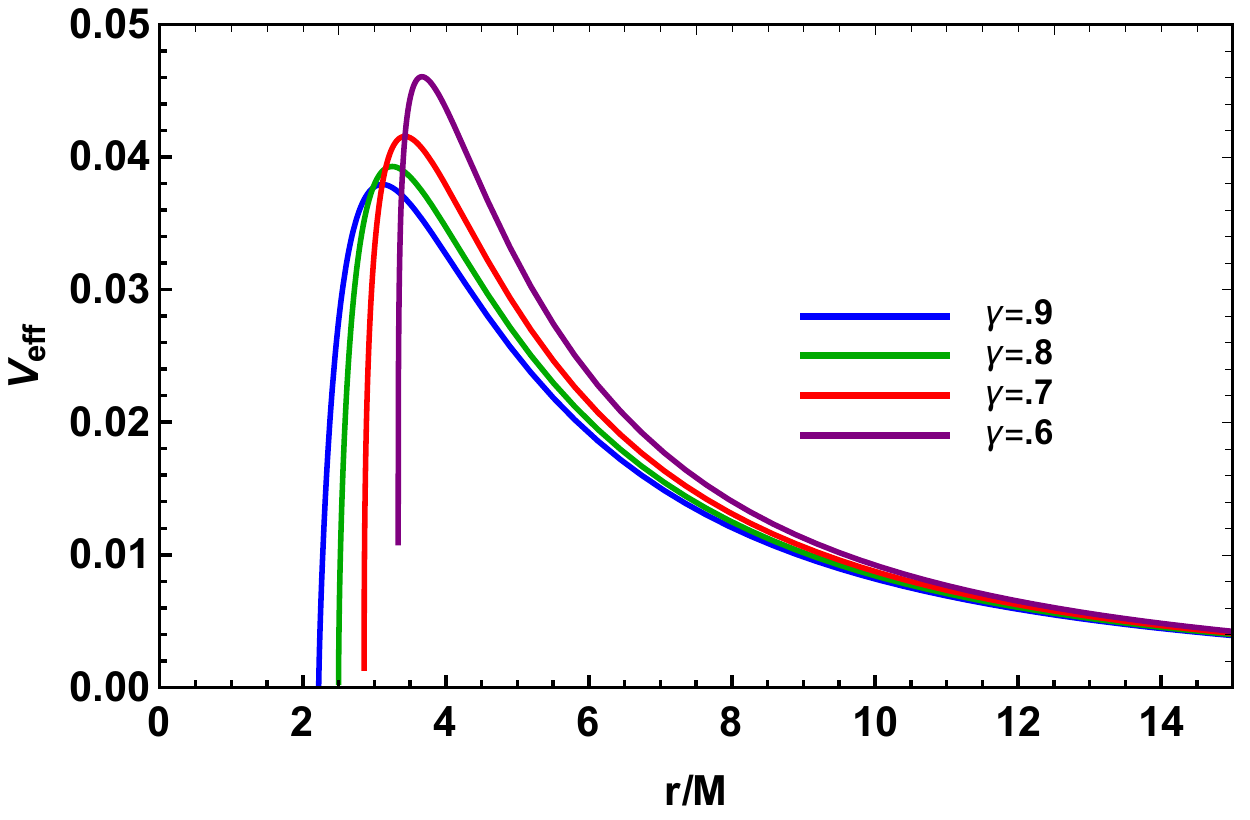}}
\subfloat[\label{f1b}]{\includegraphics[scale=0.68]{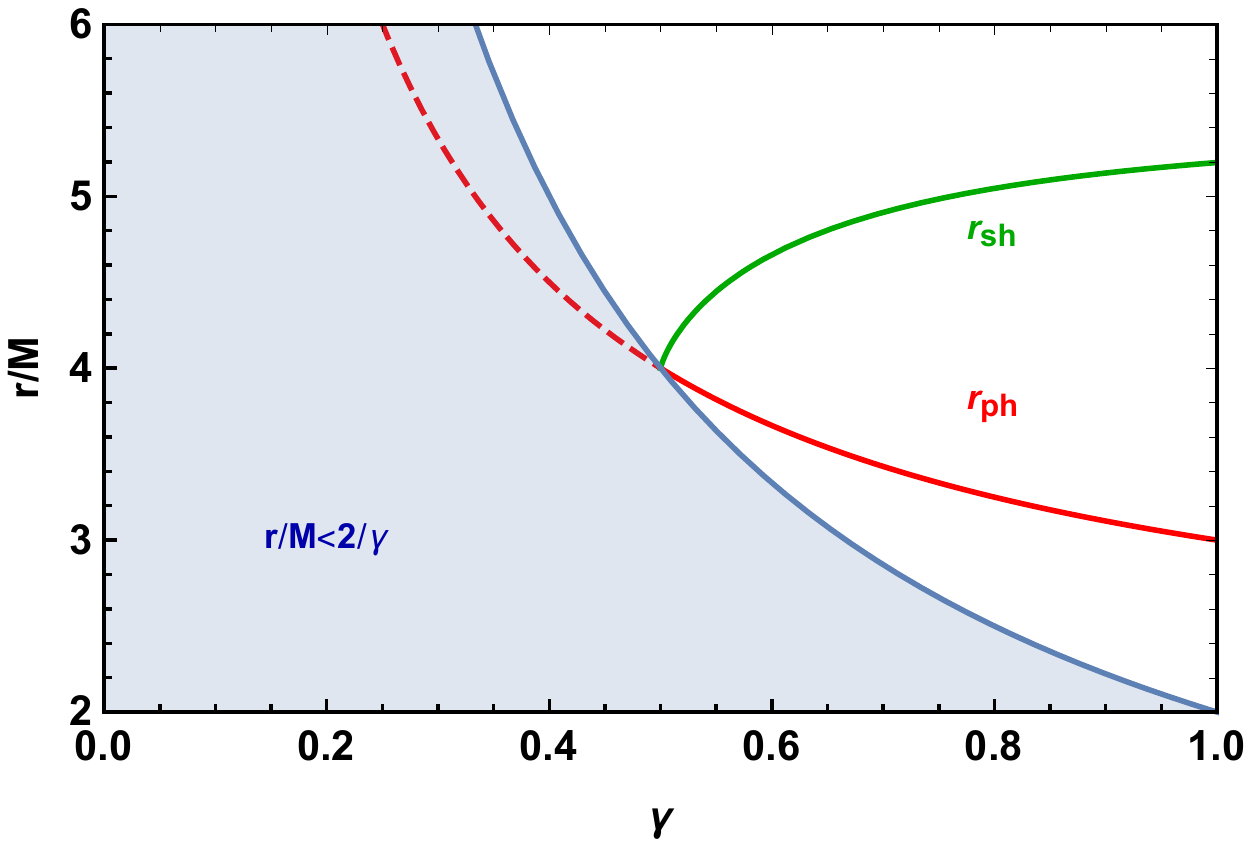}}
%\subfloat[\label{f2c}]{\includegraphics[scale=.43]{Capture.pdf}}
\caption{The above figure depicts the dependence of (a) the effective potential (b) the photon sphere $r_{ph}$ and the radius of the shadow $r_{sh}$ on the metric parameter $\gamma$. The metric has a curvature singularity at $r_c=b$ where $b=2M/\gamma$. 
The region of unphysical solutions $(r<b)$ is shaded in blue.
We note that at $\gamma=0.5$ both $r_{ph}=r_{sh}=r_c=b$. When $\gamma<0.5$, 
the photon sphere disappears since $r_{ph}<r_c$. We therefore confine ourselves in the region $r>r_c$ and $0.5\leq\gamma\leq 1$. }
\label{f1}
\end{figure}

In this section we will study the properties of the shadow given by the metric in \ref{S2_1}. Before we proceed with the discussion of the shadow, we first plot the effective potential discussed in the last section in \ref{f1a}. The figure depicts the behaviour of the effective potential with the variation of the metric parameter $\gamma$. 
As expected, the effective potential $V_{eff}$ has a maxima occuring at the photon sphere $r_{ph}$, which depends on the value of $\gamma$. 
On decreasing the scalar charge $q$ (or increasing $\gamma$), $r_{ph}$ becomes smaller along with the height of the potential.  

We can further determine 
the radius of the photon sphere and the shadow using \ref{S3_10} and \ref{28} for the metric in \ref{S2_1}.  These are given by
%%%%%%%%%%%%%%%%
\begin{flalign}
 & r_{ph}=b\left(\gamma+\dfrac{1}{2}\right)~~~~~~\rm{and}
 \label{rph-1}
\end{flalign}
%%%%%%%%%%%%%%%%
\begin{flalign}
& r_{sh}=b\left(\gamma+\dfrac{1}{2}\right)\left(\dfrac{2\gamma-1}{2\gamma+1}\right)^{\frac{1}{2}-\gamma}\label{Sh_1}
\end{flalign}
%%%%%%%%%%%%%%%%%%
respectively. In \ref{rph-1} and \ref{Sh_1} the $r_{ph}$ and $r_{sh}$ are expressed in units of $M$. In what follows we will scale the radial coordinate by the mass $M$ of the black hole, such that $r\equiv r/M$. Consequently, $b\equiv b/M$, the scalar charge $q\equiv q/M$ and the axion parameter $h\equiv h/M^2$.
From \ref{S2} we recall that $b\gamma=2$ and $0\leq\gamma\leq 1$. We note that as we decrease the value of $\gamma$ from unity, the radius of the photon sphere $r_{ph}$ increases while that of the shadow $r_{sh}$ decreases. At $\gamma=0.5$ both $r_{ph}$ and $r_{sh}$ become equal to $r_c=b$, the radius where the curvature singularity occurs. When $\gamma<0.5$, $r_{ph}<r_c$ and therefore we confine ourselves in the region $0.5\leq\gamma\leq 1$. The above discussion is illustrated in \ref{f1b}. The region of unphysical solutions $(r<b)$ is shaded in blue. As soon as the field parameter $\gamma$ approaches the critical 
value $\gamma=0.5$, physical solutions for $r_{ph}$ and $r_{sh}$ ceases to exist.

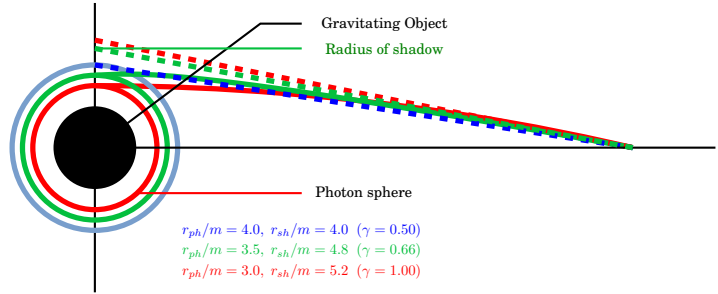
\begin{figure}[t!]
\begin{center}
\begin{tikzpicture}[thick,scale=0.55, every node/.style={transform shape}]
%\draw[flame] (0,0) circle (3cm);
\fill[black] (0,0) circle (1cm);
\draw (0,-3.5)--(0,3.5);
\draw[red,line width=.7mm] (0,0)circle (1.5cm);
\draw[red,line width=.7mm] (0,1.5) parabola (13,0); 
\draw[darkpastelblue,line width=.7mm] (0,0)circle (2cm);
\node[blue,line width=.7mm] at (5,-2) {$r_{ph}/m=4.0,~r_{sh}/m=4.0~~(\gamma=0.50)$};
\node[red,line width=.7mm] at (5,-3) {$r_{ph}/m=3.0,~r_{sh}/m=5.2~~(\gamma=1.00)$};
\node[darkpastelgreen,line width=.7mm] at (5,-2.5) {$r_{ph}/m=3.5,~r_{sh}/m=4.8~~(\gamma=0.66)$};
\draw[darkpastelgreen,line width=.7mm] (0,0)circle (1.75cm);
\draw[darkpastelgreen,line width=.7mm] (13,0).. controls (2,2) and (1,1.8)..(0,1.75);
\draw[blue,dashed,line width=.7mm] (0,2) -- (13,0);
\draw[dashed,red,line width=.7mm] (0,2.6)--(13,0);
\draw (0,0)--(15,0);
\draw[darkpastelgreen,dashed,line width=.7mm] (0,2.4)--(13,0);
\draw[] (0,0)--(4,3)--(5,3);
\draw[red](1,-1.1)--(5,-1.1);
\node[line width=.7] at (6.5,-1.1) {Photon sphere};
\draw[darkpastelgreen] (0,2.4)--(5,2.4);
\node[line width=.7] at (7,3) {Gravitating Object};
\node[line width=.7,ao(english)] at (7,2.4) {Radius of shadow};
\draw[];
\end{tikzpicture}
\caption{Diagrammatic realisation of gravitational lensing in Winicour spacetime}
\label{f2}
\end{center}
\end{figure}

At a glance, this atypical behaviour of photon sphere and shadow seems counter-intuitive
since the radius of the shadow generally increases with the radius of the photon sphere.
In this respect the behavior of the Winicour solution is quite unique. 
But one can understand this scenario with the analogy of having an optical lens system in a medium denser than air.  
Optical system in relatively denser medium bends light relatively smaller. Similarly, increasing the scalar charge $q$ is equivalent to putting the optical system in a relatively denser medium i.e. a medium with larger refractive index. Consequently, light bending is maximum in the Schwarzschild scenario compared to the situation where there is scalar charge.    
The diagrammatic realisation has been shown in \ref{f2} which clearly shows that the presence of scalar field causes lesser deflection of light compared to the Schwarzschild scenario.
%%%%%%%%%%%%%
%%%%%%%%%%%%%%%%%%%%%%%
%%%%%%%%%%%%%%%
\paragraph{}
Finally we end our discussion with a few interesting comments:
\begin{itemize}
\item We have noted from \ref{S2} that the scalar field and the Kalb-Ramond field both minimally coupled to gravity give rise to identical energy-momentum tensor.
However, in the case of the scalar field the solution of the gravitational field equations lead to an exact metric representing a naked singularity while in the other case the solution leads to a perturbative metric representing a black hole. We have explored in an earlier work \cite{Banerjee:2019xds} the dependence of the shadow radius on the axion parameter $h$ and found that a negative $h$ enhances, while a positive $h$ diminishes the shadow compared to the Schwarzschild scenario \ref{f3a}. In the Winicour solution on the other hand, the shadow decreases with decrease in $\gamma$ (or increase with the scalar charge $q$) and its radius is always less than the Schwarzschild case. This is illustrated in \ref{f3b}.

%%%%%%%%%%%%%%%%%%%%%%%%%%%%%%
\begin{figure}[t!]
\begin{center}
\subfloat[\label{f3a}]{\includegraphics[scale=0.45]{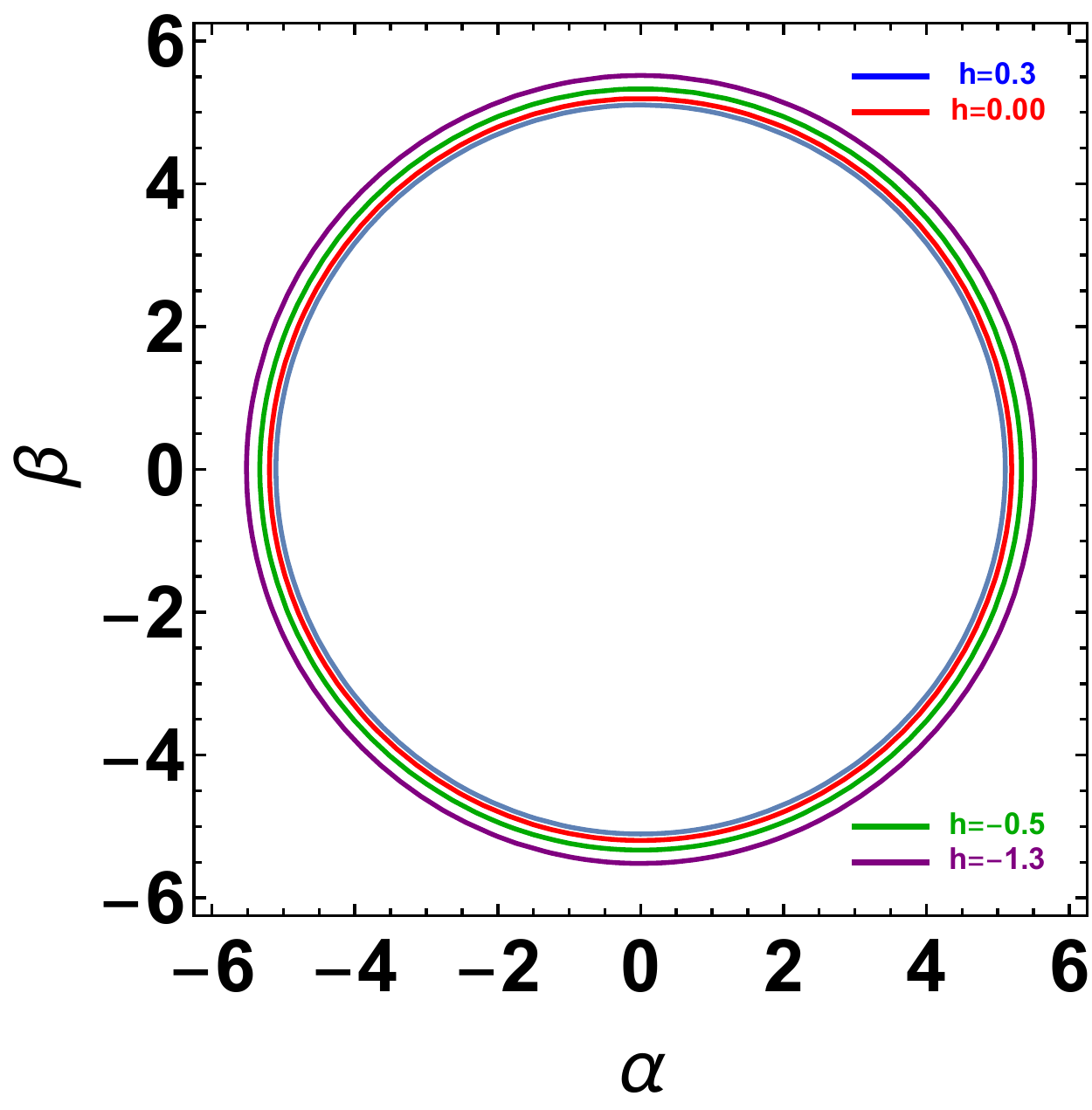}}
\subfloat[\label{f3b}]{\includegraphics[scale=0.47]{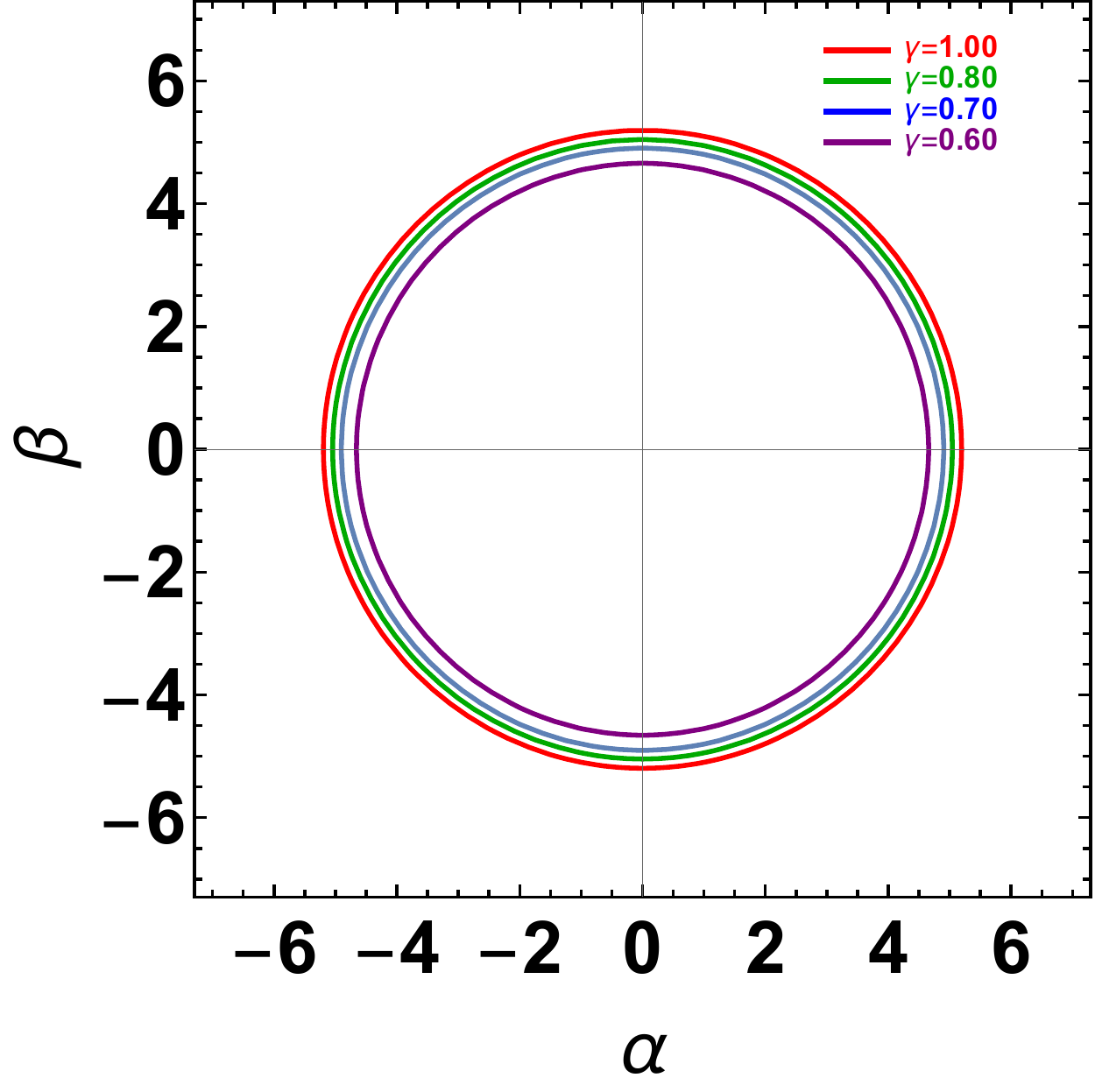}}
\caption{Radius of the shadow for (a) the perturbative axion metric and (b) the exact Winicour solution for various values of their respective metric parameters. }
\label{f3}
\end{center}
\end{figure}

\item We note from \ref{f3b} that the Schwarzschild scenario produces larger radius of the shadow compared to the ones with non-trivial scalar charge. Further, if we allow the black hole to be rotating in \gr\ (the Kerr black hole), the radius of the shadow also turns out to be smaller than the Schwarzschild scenario. It is therefore interesting to understand if a Kerr black hole can be distinguished from the ones with scalar charge from shadow related observations.

\begin{figure}[t!]
\centering
\subfloat[\label{f4a}]{\includegraphics[scale=0.68]{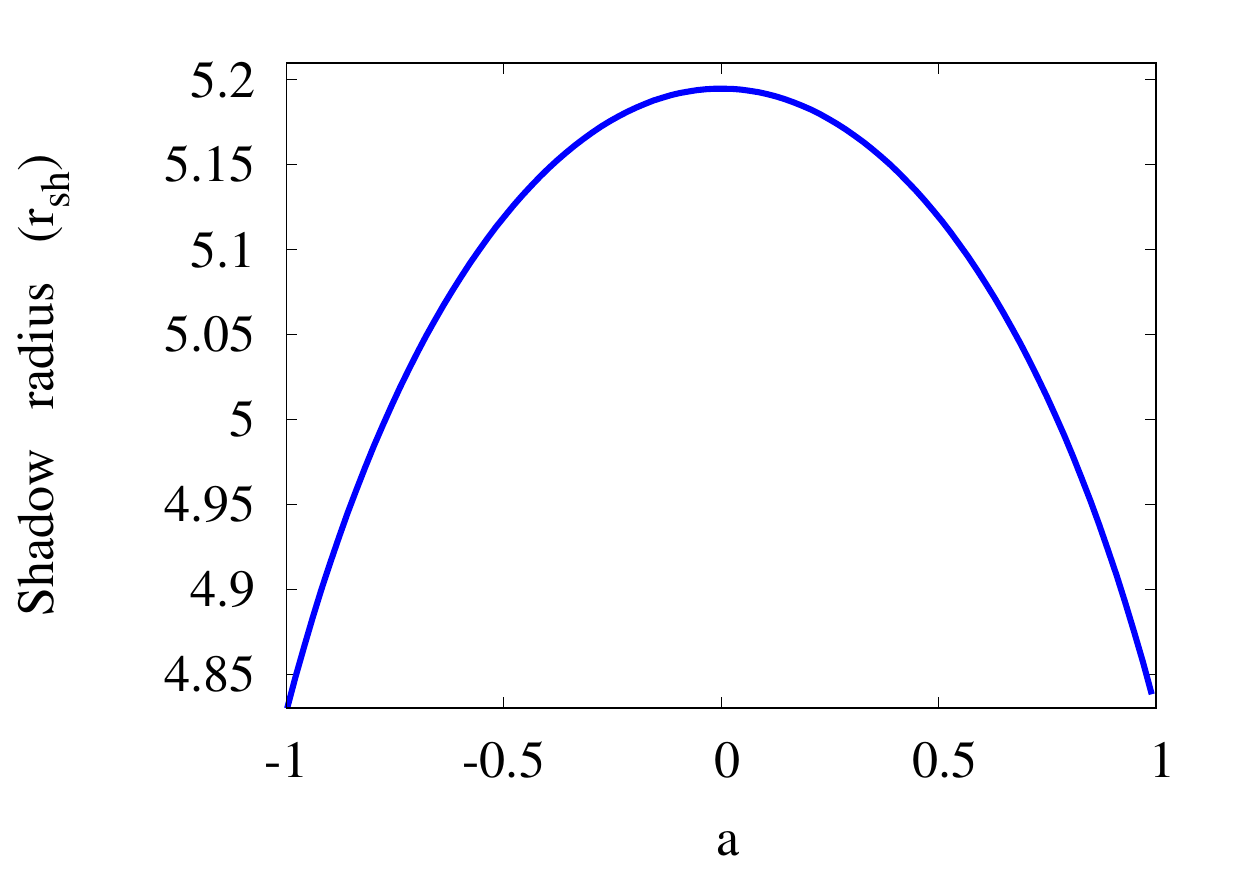}}
\subfloat[\label{f4b}]{\includegraphics[scale=0.667]{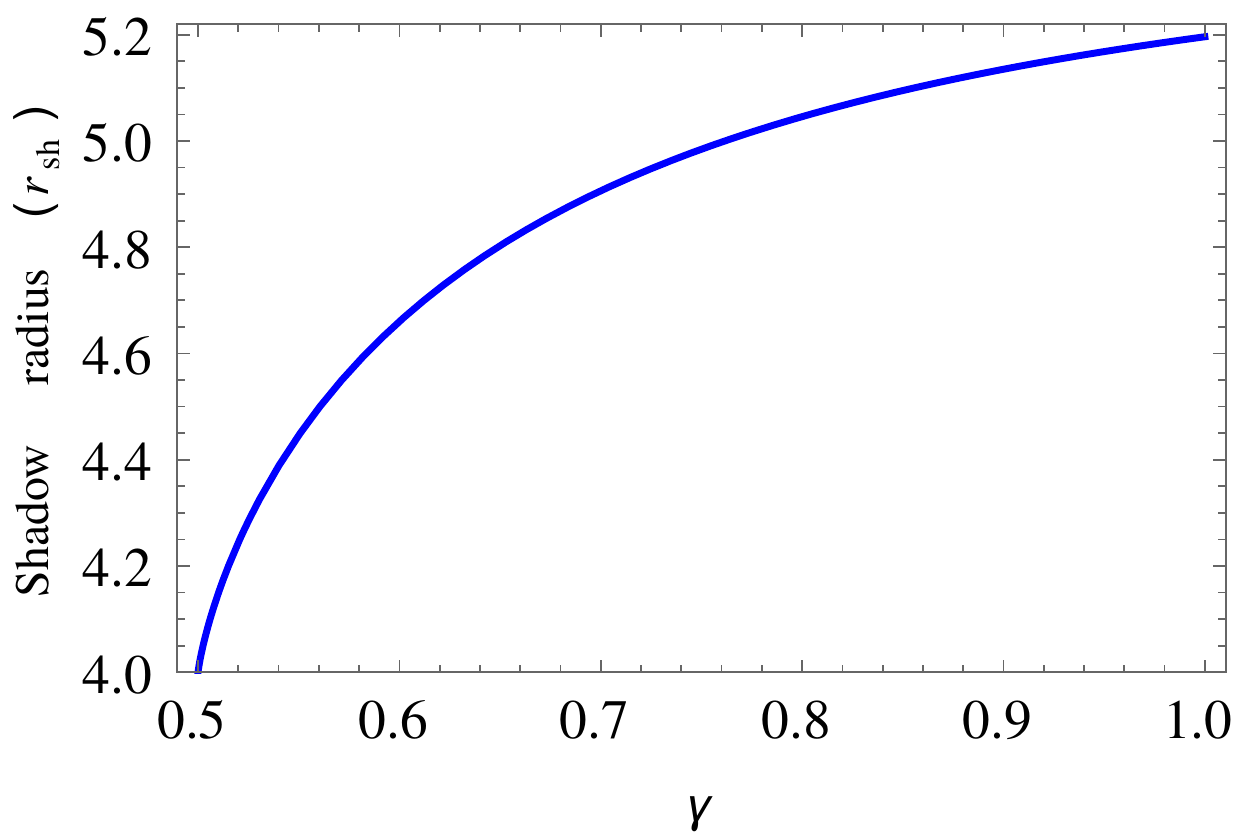}}
%\subfloat[\label{f2c}]{\includegraphics[scale=.43]{Capture.pdf}}
\caption{The figure illustrates the variation of the shadow radius with (a) the Kerr parameter $a$ (the black hole viewed at $i=0$) and (b) the metric parameter $\gamma$ of the Janis Newman Winicour spacetime.  }
\label{f4}
\end{figure}

In this context it is important to note that the spin $a$ of the black hole not only affects the size of the shadow but also its shape. This becomes pronounced at a high inclination angle as the presence of angular momentum leads to a dented shadow thereby causing a deviation from circularity in its shape. At low inclination angles this effect is less conspicuous. 
In particular, it can be shown that if a black hole is viewed face on (zero inclination angle) then the shadow is circular although the radius of the shadow depends on the black hole hole angular momentum. To elucidate this point we note that  the $x$ and $y$ coordinates of the shadow for a Kerr black hole are given by,
\begin{align}
\label{eq1}
x=-\frac{l}{sin i} ~~~~~~~~~~~~~y=\pm \sqrt{\chi+a^2cos^2i-l^2cot^2i}
\end{align}
In \ref{eq1}, $i$ refers to the inclination angle, $a$ is the dimensionless black hole spin parameter, $l=L/E$ and $\chi=C/E^2$ are the two impact parameters, such that $L$ is the specific angular momentum, $E$ is the specific energy and $C$ refers to the Carter constant. The derivation of \ref{eq1} can be found in \cite{Vries_1999,Cunha:2018acu}.
From \ref{eq1} it can be shown that when $i=0$ the contour of the shadow is given by,
\begin{align}
x^2+y^2=\chi+a^2=r^2_{sh}~~~~~~~\rm where
\label{eq2}
\end{align}

\begin{align}
\label{eq3}
\chi=-\frac{r_{ph}^3(r_{ph}^3+9r_{ph}-6r_{ph}^2-4a^2)}{a^2(r_{ph}-1)^2}
\end{align}
depends on $a$ and $r_{ph}$ \cite{Teo2003,Vries_1999,Cunha:2018acu}. The radius of the photon sphere $r_{ph}$ in turn also depends on $a$ and is given by,
\begin{align}
\begin{cases}
r_{ph}=1+\sqrt{A}[\lbrace B +\sqrt{B^2+1}\rbrace^{\frac{1}{3}}+  \lbrace B +\sqrt{B^2+1}\rbrace^{-\frac{1}{3}}]~~~~~\rm {if} ~|B| > 1\\
r_{ph}=1+2\sqrt{A}cos\big(\frac{1}{3}cos^{-1}B \big) ~~~~~~~~~~~~~~~~~~~~~~~~~~~~~~~~~~~~~~~~~~~~\rm {if} ~|B| \leq 1
\end{cases}
\label{4}
\end{align}
where
\begin{align}
A=\frac{3-a^2}{3} ~~~~~~~B=\frac{1-a^2}{A^{\frac{3}{2}}}
\label{5}
\end{align}

\ref{f4} depicts the variation of the shadow radius with the Kerr parameter $a$ (considering $i=0$) and the JNW metric parameter $\gamma$. It is clear from \ref{f4a} that for $i=0$ an increase in Kerr parameter decreases the shadow radius. Similarly, if we have a spherically symmetric black hole with scalar charge described by the JNW spacetime, the radius of the shadow diminishes with an increase in scalar charge $q$ (or decrease in $\gamma$).
However, the degree of reduction in the shadow radius is more due to the presence of scalar charge than when it is rotating ($r_{sh}\sim 4.0~R_g$ when $\gamma=0.5$ while $r_{sh}\sim 4.83 ~R_g$ when $|a|\sim 1$ compared to $r_{sh}\sim 5.196 R_g$ in the Schwarzschild scenario). This directly affects the angular diameter $\theta$ of the shadow, since 
\begin{flalign}
\tan \theta \approx \theta = \dfrac{ 2r_{sh}GM}{Dc^2} \label{eq8}
\end{flalign}
such that $M$ is the mass and $D$ is the distance of the black hole from the observer.
This result has important
implications with respect to the observed shadow of M87* which we shall discuss in the next section.

%If the mass and distance of a black hole are precisely estimated then the radius of the shadow can be used to constrain/discard various alternate gravity models. In particular, in a future observation if a black hole has precise and independent measurements of its mass and distance, casts a circular shadow irrespective of the inclination angle and the measured angular diameter turns out to be smaller than the prediction of \gr\ then the possibility that the background is described by the JNW metric gets enhanced. 

\item We could have considered the metric ansatz in \ref{S2_1} by removing the constraint $b\gamma=2M$ and kept $\gamma$ and $b$ independent. 
Such a metric ansatz is compatible with the Einstien's equations with the minimally coupled scalar field. We refer to such a metric as the generalized Janis-Newman-Winicour solution. The interesting characteristic of this generalised solution is that we do not get physical solutions for the photon sphere and the shadow for all values of the metric parameters $\gamma$ and $b$.

\begin{figure}[t!]
\begin{center}
\subfloat[\label{Fig5a}]{\includegraphics[scale=0.7]{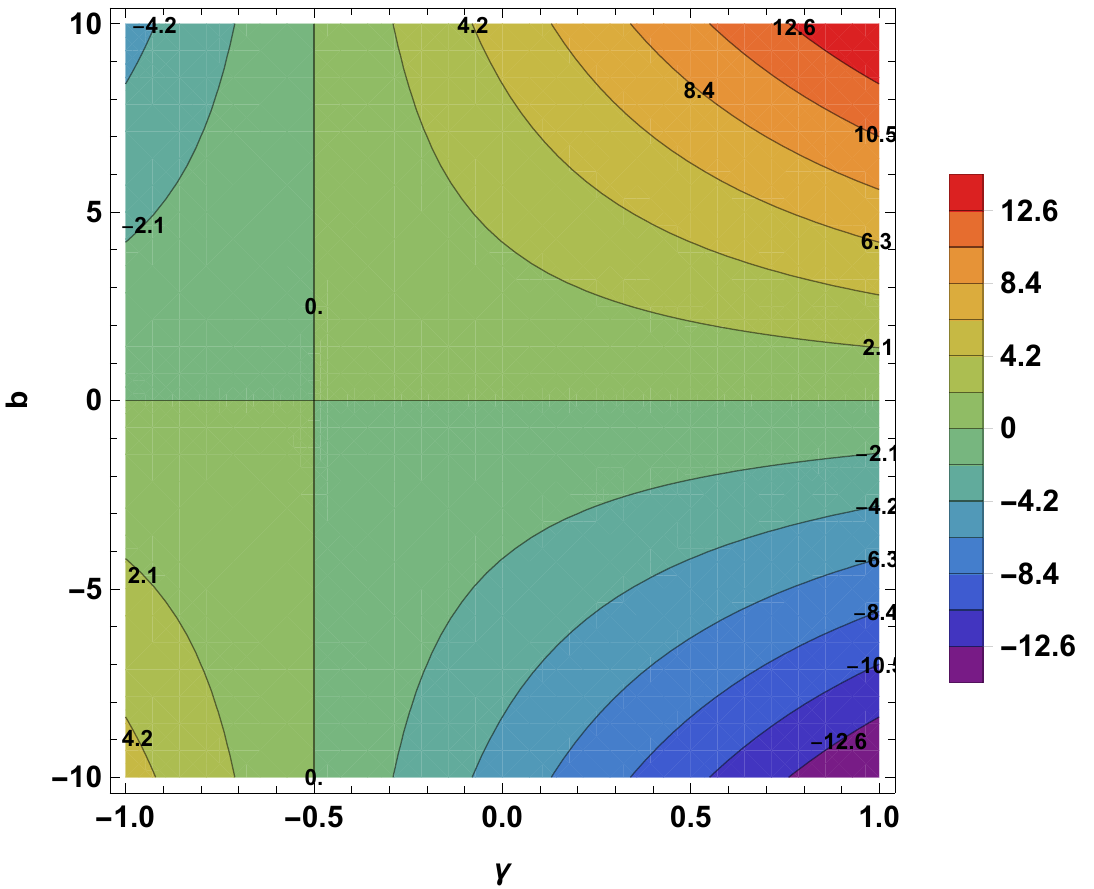}} 
\subfloat[\label{Fig5b}]{\includegraphics[scale=0.7]{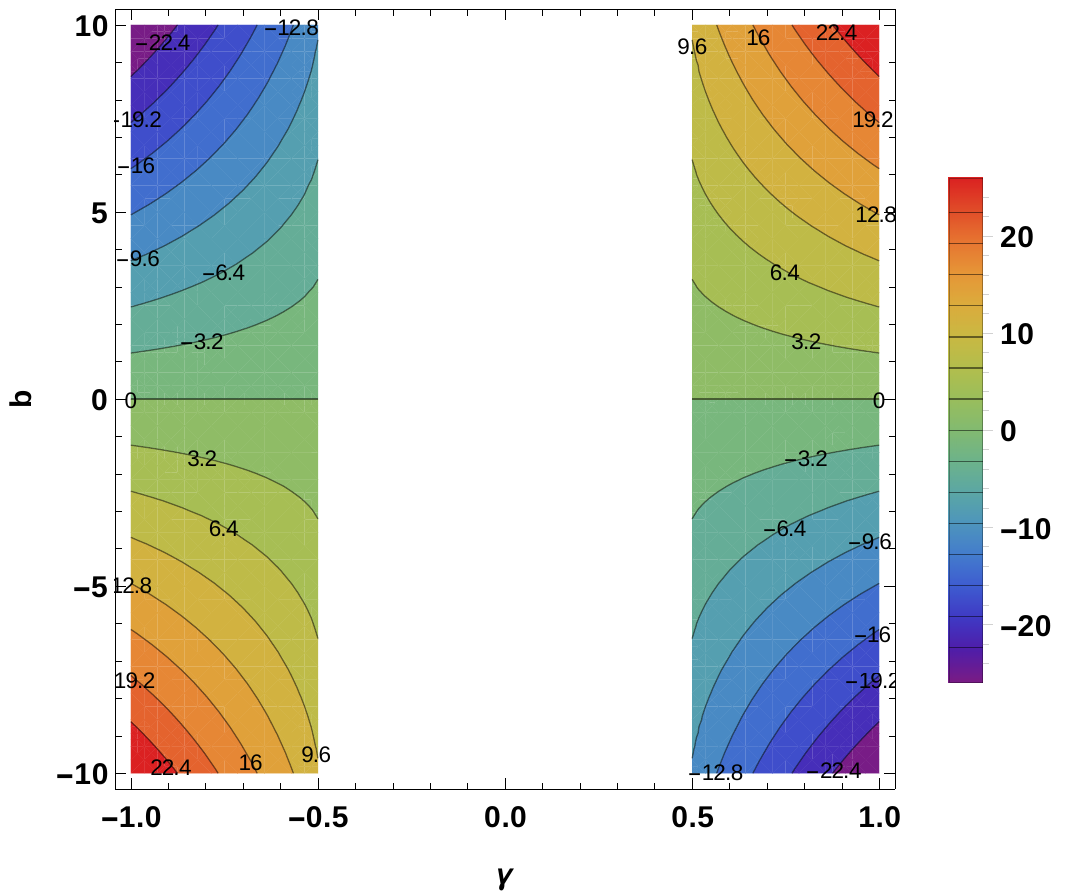}} 
\caption{The above figure represents the constant contours of the radius of the (a) photon sphere and (b) the shadow as functions of the metric parameters $\gamma$ and $b$. The shadow and the photon sphere are expressed in units of $GM/c^2$.}
\label{f5}
\end{center}
\end{figure}

In \ref{f5}, the regions $b>0 $, $\gamma<0$ and $b<0$, $\gamma>0.5$, 
produces negative radii for the photon sphere and the shadow and hence are not physically important. As discussed in \ref{S4.2} real positive solution of the shadow is achievable only if $|\gamma|>0.5$. In fact for the region $|\gamma|<0.5$, no physically realizable solution of photon sphere and shadow can be found (\ref{f1b}). Hence the observation of shadow may be possible if $b>0$, $\gamma>1/2$ or $b<0$, $\gamma<-1/2$. The second scenario where $b$ and $\gamma$ are negative is not much discussed in the literature.
However when we are in the region $b>0$ we must have $\gamma b=2$,  so that we can reproduce the Schwarzschild limit for the gravitating object. This particular 
case when $b>0$ is widely known as Janis-Newman-Winicour solution. Hence in this region, our two 
parameter solution reduces to one parameter solution discussed earlier. For the particular case when $\gamma=1$, 
the solution in \ref{S2_1} represents the Schwarzschild solution.

%\section{A comparison of shadow behaviour for exact solution and perturbative solution}

\end{itemize}

\subsection{Comparison with the observed shadow of M87*}
We have noted in the last section that a Kerr black hole casts a non-circular shadow only if it is viewed at a high inclination angle. On the other hand, if a black hole casts a circular shadow despite being viewed at high inclination angle, then it implies that the background spacetime is spherically symmetric. Further, if the black hole has precise and independent measurements of mass and distance, then the size of the observed angular diameter can be used to compare between various background spacetimes. Since the angular diameter directly depends on the shadow radius (\ref{eq8}), an observed angular diameter smaller than the Schwarzschild scenario might favor the JNW spacetime.
Therefore in a future observation, if a black hole is viewed at high inclination angle and has precise and independent measurements of its mass and distance, then the shape and size of the shadow can be a useful tool to probe the background spacetime. In this way the degeneracy between the effect of spin and $\gamma$ can be broken, although we need to wait for future observations for this. 

At present, only the angular diameter of M87*, the supermassive black hole at the centre of the galaxy M87, has been measured which corresponds to $42\pm 3\mu as$. The object exhibits a strong jet and the angle of inclination is taken to be $17^\circ$ which the jet axis makes to the line of sight. This is in ageement with the nearly circular shadow observed in M87* with deviation from circularity $\Delta C\leq 10\%$ \cite{Akiyama:2019cqa}. 
Based on stellar population measurements, the distance of M87* is reported to be $D=(16.8\pm0.8)~\textrm{Mpc}$ \cite{Blakeslee:2009tc,Bird:2010rd,Cantiello:2018ffy}. The mass of the source is constrained to be $M\sim 6.2^{+1.1}_{-0.5}\times 10^9 M_\odot$ \cite{Gebhardt:2011yw} from stellar dynamics observations while $M\sim 3.5^{+0.9}_{-0.3}\times 10^9 M_\odot$ \cite{Walsh:2013uua} from gas dynamics studies. Note that these are independent mass estimations of the object which does not depend on observations related to its shadow. From the measured angular diameter of the shadow of M87* and assuming general relativity, the EHT Collaboration has reported the mass of the object to be $M=(6.5\pm 0.7)\times 10^{9}~M_{\odot}$ \cite{Akiyama:2019cqa,Akiyama:2019fyp,Akiyama:2019eap}. Therefore, this mass measurement should not be used to constrain the background metric from shadow related observations.

\begin{table}[t!]
\caption{Variation of angular diameter of M87* with Kerr parameter $a$, JNW metric parameter $\gamma$ and black hole mass $M$. The distance of the source is assumed to be $D=16.8~\textrm{Mpc}$ while computing the angular diameter.}

\vskip 5mm
\begin{tabular}{|c|c|c|c|c|c|c|c|}
\hline
\multirow{3}{*}{Serial No.} & \multirow{3}{*}{\begin{tabular}[c]{@{}c@{}}Mass\\ (In units of $10^{9}M_{\odot}$)\end{tabular}} & \multicolumn{6}{c|}{Angular diameter (in $\mu as$)}    \\ \cline{3-8} 
                            &                                                                                                           & \multicolumn{3}{c|}{Kerr metric (i=0)} & \multicolumn{3}{c|}{JNW metric} \\ \cline{3-8} 
                            &                                                                                                           & {\cellcolor[HTML]{96FFFB}\color{blue} $a=1.0$}      & {\cellcolor[HTML]{96FFFB}\color{blue} $a=0.5$}      & {\cellcolor[HTML]{96FFFB}\color{blue} $a=0.0$}     & {\cellcolor[HTML]{96FFFB}\color{blue} $\gamma=0.5$}     & {\cellcolor[HTML]{96FFFB}\color{blue} $\gamma=0.9$}    &{\cellcolor[HTML]{96FFFB}\color{blue} $\gamma=1.0$}    \\ \hline

\multirow{3}{*}{1}          & 3.5+0.9                                                                                                   & 25.061   &  26.579  & 26.972    & 20.763   & 26.655   &   26.972\\ \cline{2-8} 
                            & 3.5                                                                                                       & 19.935   & 21.143   & 21.455    & 16.516     & 21.203     & 21.455  \\ \cline{2-8} 
                            & 3.5-0.3                                                                                                   & 18.227  & 19.331  & 19.616  & 15.101    &  19.385    &   19.616\\ \hline
                            
\multirow{3}{*}{2}          & 6.2+1.1                                                                                                   & 41.579   & 44.098  & 44.748   & 34.448    &  44.223    &   44.748     \\ \cline{2-8} 
                            & 6.2                                                                                                       & 35.314   & 37.453   & 38.006  &  29.258   &  37.559       &  38.006  \\ \cline{2-8} 
                            & 6.2-0.5                                                                                                   & 32.466     & 34.432    & 34.941  & 26.898    & 34.53      & 34.941   \\ \hline                            
                            
 \multirow{3}{*}{3}          & 6.5+0.7                                                                                                   & 41.009   & 43.494     & 44.135   & 33.977     & 43.617     &   44.135     \\ \cline{2-8} 
                            & 6.5                                                                                                       & 37.023   &  39.266   & 39.845   &  30.673   &  39.377     &   39.845    \\ \cline{2-8} 
                            & 6.5-0.7                                                                                                   & 33.035   & 35.037    & 35.554   &  27.37   &   35.136   &  35.554   \\ \hline

\end{tabular}
\label{T1}
\end{table}

The above discussion reveals that the independent mass measurements of M87* (based on stellar and gas dynamics studies) differ quite substantially. 
Further, from \ref{eq8} it is clear that the angular diameter is highly sensitive to the estimated magnitude of $M$.
In \ref{T1} the angular diameter of M87* is estimated in both the Kerr and the JNW background assuming the different mass measurements of the object (the mass reported by the EHT Collaboration is also given for completeness), while the distance is taken to be $D=16.8$ Mpc. From the table it is clear that the variation in mass affects the angular diameter much more than a change in the background spacetime (\ref{T1}). Moreover, for higher masses, a change in $\gamma$ affects the angular diameter much more than a modification in the Kerr parameter. If independent mass estimations are not available then the angular diameter of the shadow can be used to determine the mass assuming a given background metric (as done by the EHT collaboration). In such a scenario, however, one cannot constrain the background from the angular diameter.
Alternatively, without independent mass measurements the degeneracy between the mass and the background spacetime cannot be broken from the observed angular diameter of the shadow.

\begin{figure}[t!]
\centering
\subfloat[\label{f6a}]{\includegraphics[scale=0.68]{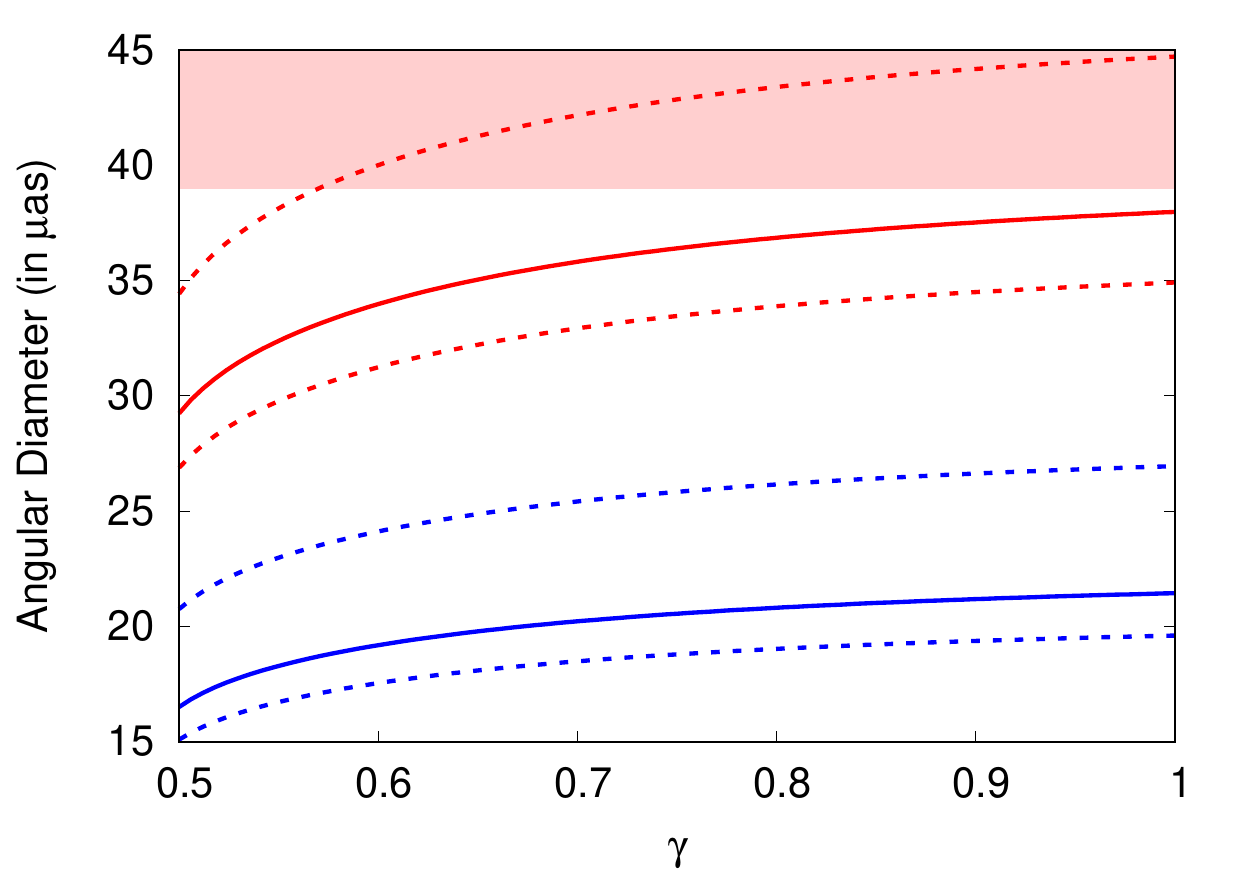}}
\subfloat[\label{f6b}]{\includegraphics[scale=0.68]{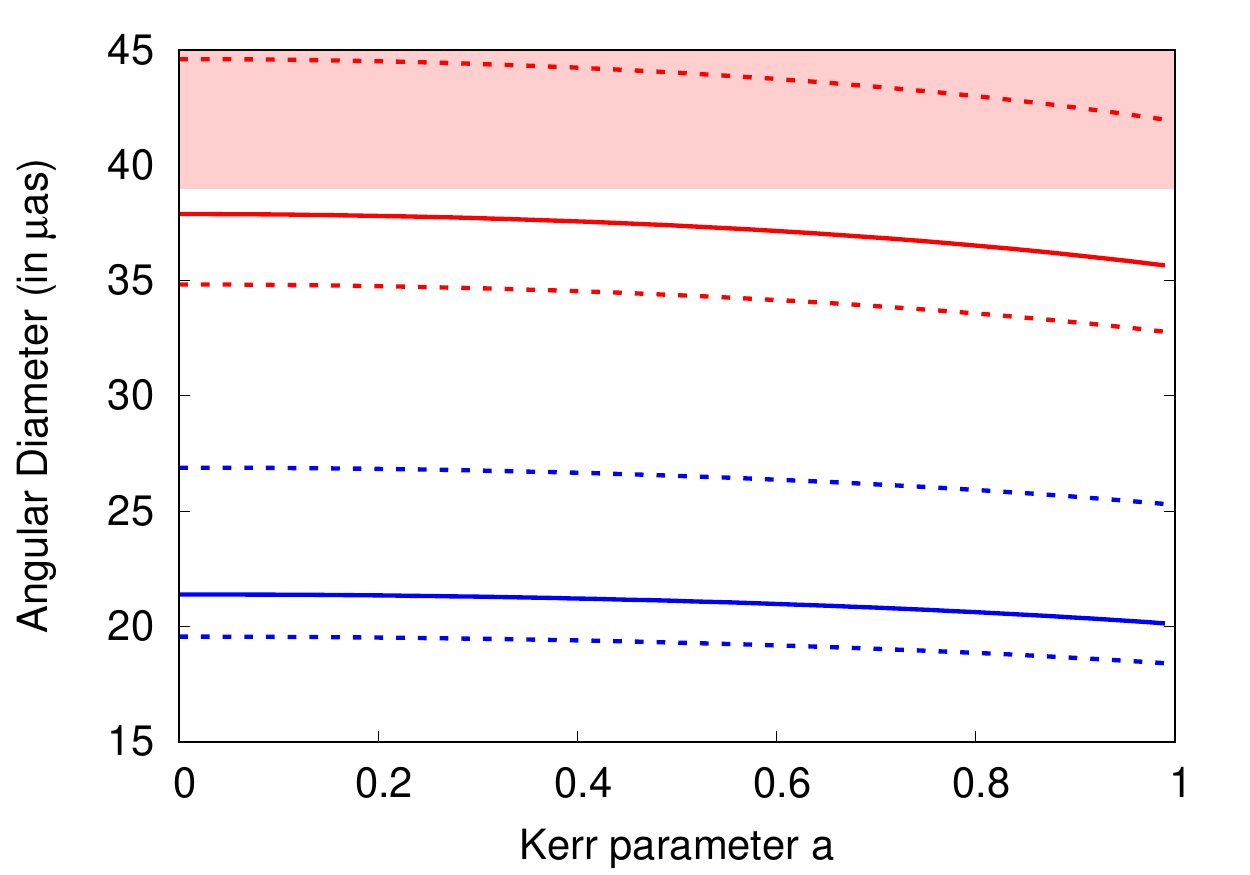}}
%\subfloat[\label{f2c}]{\includegraphics[scale=.43]{Capture.pdf}}
\caption{The figure illustrates the variation of the angular diameter with (a) the JNW metric parameter $\gamma$ and (b) the Kerr parameter $a$. In each of the figures, the red curves represent the angular diameter calculated with $M\sim 6.2^{+1.1}_{-0.5}\times 10^9 M_\odot$, while the blue curves are plotted assuming $M\sim 3.5^{+0.9}_{-0.3}\times 10^9 M_\odot$. The dashed curves in both the figues are plotted assuming the error bars in the masses about the central value.
The pink shaded region represents the observed angular diameter of $42\pm 3\mu as$.}
\label{f6}
\end{figure}

Although a black hole viewed at a high inclination angle can probe the background spacetime better, the present observation of M87* (viewed at $i=17^\circ$) can be used to throw some light on the mass of M87* and the viability of the JNW background. This is due to the greater reduction in the shadow radius in the JNW background compared to \gr\ (\ref{f4} and \ref{T1}). In \ref{f6a} we plot the variation in the angular diameter of the shadow with the JNW metric parameter $\gamma$ assuming $M\sim 6.2^{+1.1}_{-0.5}\times 10^9 M_\odot$ (the red curves) and $M\sim 3.5^{+0.9}_{-0.3}\times 10^9 M_\odot$ (the blue curves), which are the two independent mass measurements of the object. For comparison with \gr, the angular diameter of the shadow is also plotted against the Kerr parameter using the aforesaid masses in \ref{f6b}. In both the cases the distance is taken to be $D=16.8$ Mpc.
The angular diameter in \ref{f6} when plotted with the central value of the mass is denoted by the solid curves while the dashed curves represent the theoretical angular diameter plotted with the error bars in the masses. 
The pink shaded region in \ref{f6} denotes the observed angular diameter of $42\pm 3\mu as$. It is important to note that since $i=17^\circ$ for M87*, the shadow is not exactly circular for the Kerr black hole but elongated along the y-axis. We consider the major axis (the maximum distance between two points on the circumference of the shadow \cite{Wei:2019pjf,Banerjee:2019nnj}) as the shadow diameter while computing the angular diameter of the shadow. 

From \ref{f6} it is clear that if $M\sim 3.5^{+0.9}_{-0.3}\times 10^9 M_\odot$ is considered, then the observed angular diameter cannot be reproduced by merely changing the metric parameters. This mass estimation is therefore not favored by the observed shadow of M87*. In fact, even $M$ as high as $6.2\times 10^9 M_\odot$ cannot explain the observed angular diameter either in \gr\ or in the JNW background. This may be a plausible reason why the mass of M87* estimated by the EHT Collaboration (which is based on \gr) is greater than both the previous estimates. Also a background metric which inherently enhances the shadow radius compared to the Kerr scenario explains the observed angular diameter of M87* better than \gr\ (e.g. the braneworld scenario \cite{Banerjee:2019nnj}).  
In the present situation, it is difficult to break the degeneracy between the presence of the black hole spin and the scalar charge if a slightly higher mass is considered within the allowed range. For example, if $M\sim 6.5 \times 10^9 M_\odot$ is used to evaluate the theoretical angular diameter then $\gamma\geq 0.85$ and $|a|\leq 0.6$ can both reproduce the observed angular diameter of M87* within the error bars. In such a scenario
it is difficult to distinguish between the JNW scalar charge and the spin of the black hole from the image of M87*.

However, it is clear from \ref{f6a} that $\gamma=1$ (the Schwarzschild scenario) explains the observed shadow for most of the allowed values of $M$. In this sense the Schawarzschild scenario is more favored than the JNW spacetime and if $\gamma<0.57$ then  even $M\sim 7.3 \times 10^9 M_\odot$ cannot address the observation. Therefore such extreme values of $\gamma$ are ruled out by the present observation of the EHT collaboration.   
Similarly, $a=0$ covers the maximum range of observed angular diameter in \ref{f6b} given the allowed values of the estimated mass of the black hole.

However, M87* also exhibits a powerful jet with the jet power $P_{jet}\geq 10^{42} \rm erg ~s^{-1}$ and if one is confined to \gr\ then at least $|a| \sim 0.5$ is required to explain the jet power \cite{Akiyama:2019fyp}. This estimate of $a$ is also consistent with the shadow related observations if $ 6.5 \times 10^9 M_\odot \leq M \leq 7.3 \times 10^9 M_\odot$ is taken to compute the theoretical angular diameter (\ref{f6b}). Since spin plays a significant role in powering the jet, the Kerr scenario is more favored compared to the JNW spacetime if one has to also explain the observed jet power of M87*. \\
The above discussion therefore elucidates that \gr\ explains the observed angular diameter and the jet power of M87* better than the JNW background. However, it is important to note that while deriving the shadow radius we had implicitly assumed that the surrounding medium is optically thin such that the effect of the metric dominates the observed image. This may not be true and if the surrounding medium is optically thick then from the image of the surrounding accretion disk it is difficult to distinguish the JNW spacetime from the Schwarzschild metric \cite{Gyulchev:2019tvk}.
%\newpage

\section{Accretion around the Janis-Newman-Winicour spacetime}
\label{S3}
In this section we investigate the properties of the electromagnetic emission from the accretion disk in the Janis-Newman-Winicour spacetime. The continuum spectrum emitted by the accretion disk depends not only on the nature of the background metric but also on the properties of the accretion flow. 
We assume the Novikov-Thorne model \cite{1973blho,Page:1974he} for the accretion disk where the disk is considered to be geometrically thin and optically thick. Accretion takes place chiefly along the equatorial plane such that the accreting particles have large azimuthal velocity $v_\phi$ with negligible radial velocity $v_r$ and even smaller vertical velocity $v_z$. The presence of viscosity in the system endows the accreting matter a small radial velocity which enables it to inspiral and fall into the central compact object. Within the domain of the Novikov-Thorne model the accreting matter has practically negligible $v_z$ and hence the Novikov-Thorne accretion disk harbors `no outflows'. As the accreting matter inspirals, they lose gravitational potential energy which gets converted into electromagnetic radiation. This radiation interacts very effectively with the accreting matter and almost all of it is radiated out from the system and no heat is trapped with the accretion flow. A temperature gradient exists within the disk such that the inner disk is much hotter compared to the outer disk. Since matter and radiation interacts very efficiently, every annulus of the disk emits a black body commensurate with the temperature of the disk. The integrated emission from the disk is therefore a multi-temperature black body radiation. With these assumptions of the `Novikov-Thorne' model the flux from the accretion disk assumes an analytic form,
 \begin{align}
 \label{10}
F=\frac{\dot{M}_0}{4\pi\sqrt{-g}}\tilde{f}
\end{align}
where,
\begin{align}
\label{11}
\tilde{f}=-\frac{\Omega_{,r}}{(E-\Omega L)^2}\left[EL-E_{ms}L_{ms}-2\int_{r_{ms}}^r LE_{,r^\prime}dr^\prime \right]
\end{align}
where, $\Omega$, $E$ and $L$ are the angular velocity, specific energy and specific angular momentum of the accreting particle at the radial distance $r$. For a spherical symmetric metric, these can be expressed in terms of the metric parameters as,
\begin{align}
\label{12}
\Omega=\frac{d\phi}{dt}=\frac{ \sqrt{-\left\lbrace g_{\phi\phi,r}\right\rbrace \left\lbrace g_{tt,r}\right\rbrace}}{g_{\phi\phi,r}}
\end{align}

\begin{align}
\label{13}
E=-u_t=\frac{-g_{tt}}{\sqrt{-g_{tt}-\Omega^2 g_{\phi\phi}}} 
\end{align}
and
\begin{align}
\label{14}
L=u_\phi=\frac{\Omega g_{\phi\phi}}{\sqrt{-g_{tt}-\Omega^2 g_{\phi\phi}}}
\end{align}
$E_{ms}$ and $L_{ms}$ refer to the energy and angular momentum of the test particle at the marginally stable circular orbit. For a detailed discussion on the Novikov-Thorne model and a derivation of the expression for flux one is referred to \cite{1973blho,Page:1974he,Banerjee:2019sae}.

Since the photon emits a Planck spectrum at every radius, the peak temperature is given by $T(r)=\left(\tilde{F}(r)/\sigma\right)^{1/4}$ where $ \tilde{F}(r)=F(r)c^6/(G^2M^2)$ is the flux given in \ref{10} obtained after bringing back the fundamental constants and 
$\sigma$ is the Stefan Boltzmann constant. 

The luminosity from the thin accretion disk is obtained by integrating the Planck function $B_\nu(T(r))$ over the disk surface at the observed frequency $\nu$, such that,
\begin{align}
L_\nu&=8\pi^2 r_g^2\cos i  \int_{r_{\rm ms}}^{r_{\rm out}}\sqrt{-g} B_{\nu}(T(r)) dr ~~~{\rm and}\label{15}  \\
B_\nu (T)&=\frac{2h\nu^3/c^2}{{\rm exp}\left(\frac{h\nu}{z_g kT}\right)-1} \label{16}
\end{align}
In \ref{15}, $i$ refers to the inclination angle of the disk to the line of sight, $r_g=GM/c^2$ denotes the gravitational radius, and $z_g$ denotes the
gravitational redshift factor given by,
\begin{align}
z_g=E\frac{\sqrt{-g_{tt}-\Omega^2 g_{\phi\phi}}}{E-\Omega L} \label{17}
\end{align} 
The red-shift factor takes care of the modification induced in the photon frequency while travelling from the emitting material to the observer \cite{Ayzenberg:2017ufk}. 

Note that the theoretical spectrum depends chiefly on the $g_{tt}$ component of the metric while the $g_{rr}$ component and the volume factor is required only through the determinant of the metric (see \ref{10}, \ref{15}) \cite{Banerjee:2017npv}.

\begin{figure}[t!]
\centering
\hbox{\hspace{8.8em}
\includegraphics[scale=0.72]{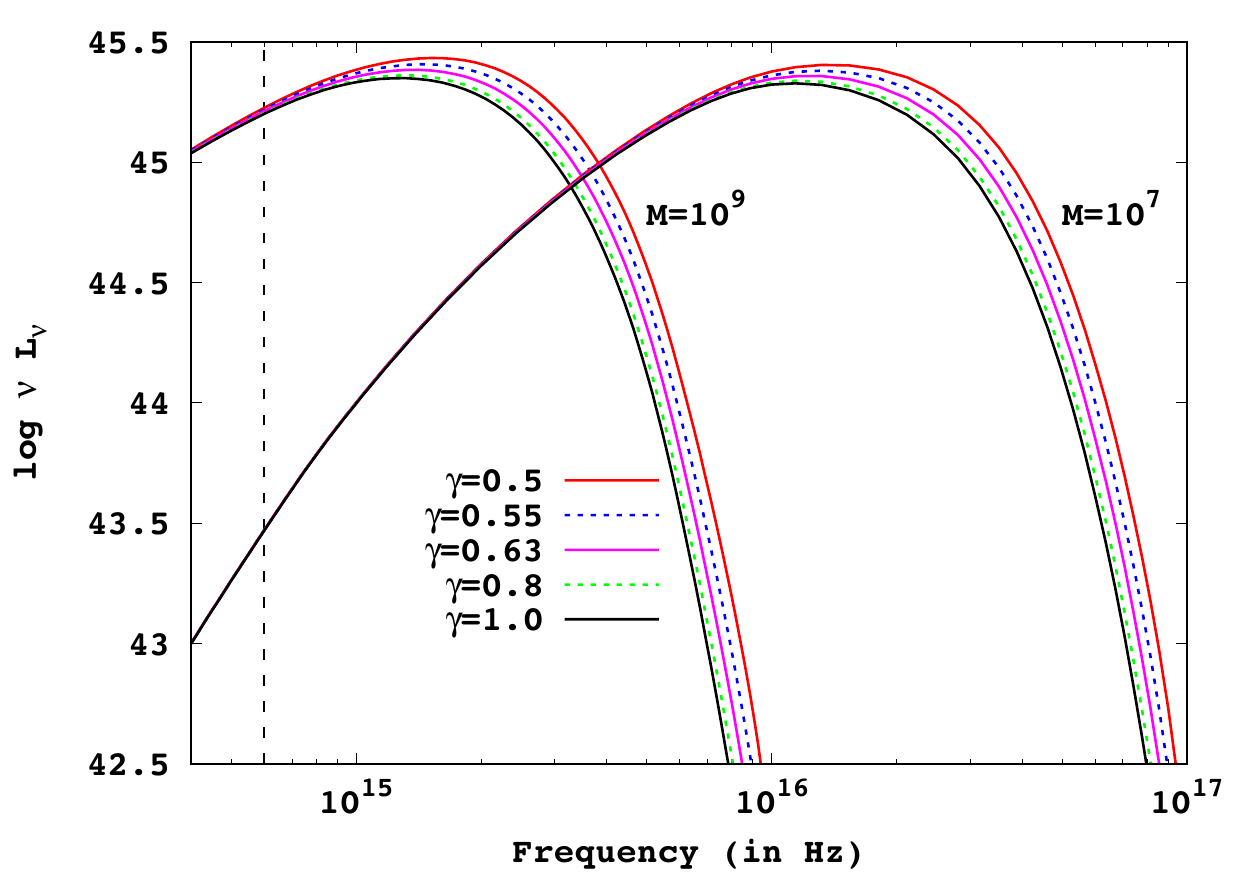}}
\caption{The above figure illustrates variation of the theoretically derived  luminosity from the accretion disk with frequency for various values of $\gamma$. The background is given by \ref{S2_1}. The luminosity decreases with increasing $\gamma$ and is minimum in the general relativistic scenario where $\gamma=1$ (the Schwarzschild scenario). The representative masses of the black hole are taken to be $M=10^9 M_\odot$ and $M=10^7 M_\odot$.
The accretion rate assumed is $1 M_{\odot}\textrm{yr}^{-1}$ and $\cos i$ is taken to be $0.8$.}
\label{Fig_01}
\end{figure}

The dependence of the theoretical spectrum from the accretion disk on the metric parameter $\gamma$ is illustrated in  \ref{Fig_01}. We note that the presence of the scalar charge enhances the luminosity from the accretion disk for both the black hole masses (The Schwarzschild scenario is represented by the black solid line) \cite{Gyulchev:2019tvk}. Since the accretion rate in \ref{10} is expressed in Eddington units, the peak temperature $T(r)\propto M^{-1/4}$ \cite{2002apa..book.....F,Davis:2010uq,Banerjee:2019sae}. Therefore, the maximum luminosity from the accretion disk around a higher mass black hole peaks at a lower frequency.

In the next section we will estimate the observationally favored value of $\gamma$ by comparing the theoretically calculated luminosity with the observed luminosity of  Palomar Green quasars.

\subsection{Observational sample}
\label{S3-1}
In this section, we compute the theoretical estimates of optical luminosity of a sample of Palomar Green quasars considered in \cite{Schmidt:1983hr,Davis:2010uq} and compare these with the corresponding observed values. 
The masses of these quasars have been independently estimated using the method of reverberation mapping \cite{Kaspi:1999pz,Kaspi:2005wx,Boroson:1992cf,Peterson:2004nu}. For a sub-sample of thirteen quasars \cite{Davis:2010uq}, the masses are also reported by $M-\sigma$ method \cite{Ferrarese:2000se,Gebhardt:2000fk,Tremaine:2002js}. 
Using observed data in the optical \cite{1987ApJS...63..615N}, UV \cite{Baskin:2004wn}, far-UV \cite{Scott:2004sv}, and soft X-ray \cite{Brandt:1999cm}, the bolometric luminosities of these quasars have been estimated \cite{Davis:2010uq}.

We calculate optical luminosity $L_{opt}\equiv \nu L_\nu$ at the wavelength 4861\AA\ \cite{Davis:2010uq} for comparison with observations. For quasars the theoretical emission from the accretion disk peaks in the far-UV/extreme UV (FUV/EUV) part of the spectrum, if the Novikov-Thorne thin disk model is considered. In the observational front on the other hand, the UV region of the spectral energy distribution (SED) is not entirely contributed by the accretion disk but some physical mechanism (e.g. advection, a Comptonizing coronae, etc.) redistributes the UV flux to the X-ray frequencies \cite{Davis:2010uq}. Therefore, although the effect of the background metric becomes most pronounced in the UV domain for quasars, extracting the effect of the metric from UV observations become difficult due to the contamination in the UV emission from components other than the accretion disk. Moreover, the error in bolometric luminosity receives maximum contribution from the far-UV extrapolation since the uncertainty in the UV luminosity far exceeds other sources of error (e.g., optical or X-ray variability) \cite{Davis:2010uq}. Therefore, a comparison of the theoretically derived UV luminosities (where the spectrum peaks) with the observed UV luminosities, might lead to erroneous conclusions regarding the background spacetime, and hence we dwell in the optical domain.

We have already discussed in the previous section that the maximum disk luminosity of a lower mass black hole peaks at a higher frequency. Therefore, the peak emission from the accretion disk of a $10^9 M_\odot$ black hole is closer to $4861\AA$, the wavelength at which the analysis is done. Since the peak of the disk emission occurs very close to the marginally stable circular orbit (msco), the emission at $4861\AA$ comes from an inner part of the disk (closer to the msco) for a $10^9 M_\odot$ black hole compared to a $10^7 M_\odot$ black hole. The wavelength $4861\AA$ corresponds to a frequency $\sim 6\times 10^{14}$ Hz and is depicted with the dashed black vertical line in \ref{Fig_01}.
Therefore, for black holes with $M\sim 10^9 M_\odot$ the effect of the metric on the emission at $4861\AA$ will be more pronounced. This motivates us to consider only the quasars with $M\ge 10^{9} M_\odot$ of the sample reported in \cite{Davis:2010uq}. It turns out that out of eighty quasars discussed in \cite{Davis:2010uq}, eleven quasars have a mass greater than a billion solar masses. We will consider only these quasars in this work.

Since quasars are not expected to be edge-on systems the inclination angle $i$ is generally believed to lie between $\cos i \in \left(0.5,1\right)$. In this work we adopt
a typical value of $\cos i \sim 0.8$ in our analysis \cite{Davis:2010uq,Wu:2013zqa}.
It turns out that for non-rotating black holes the error (e.g., reduced $\chi^2$, Nash-Sutcliffe efficiency, index of agreement etc.) between the theoretical and observed luminosities get minimized when cosi lies between $0.77- 0.82$ \cite{Banerjee:2017hzw}.
Moreover, Piotrovich et al. \cite{2017Ap&SS.362..231P} estimated the inclination angles of some of the quasars in our sample which turns out to be consistent with our choice.

The accretion rates of the quasars are reported in \cite{Davis:2010uq}. 
The accretion rates in \cite{Davis:2010uq} are estimated based on a stellar-atmosphere-like model (referred to as TLUSTY models) with black hole spin $a/M= 0.9$. However, if a blackbody model with spin $a/M= 0$ is used, then for larger $M$ the accretion rates are expected to exhibit a maximum increase by $40\%$ while for smaller $M$ the accretion rates tend to be smaller by $20\%$ compared to the accretion rates reported in \cite{Davis:2010uq}. 
In order to take this factor into account we vary the accretion rates between $80\%$ to $140\%$ of the reported accretion rates \cite{Davis:2010uq} for each quasar in the sub-sample (with $M\geq 10^9 M_\odot$), while performing the error estimations.

\subsection{Numerical Analysis and error estimators}
In this section we compute several error estimators which will enable us to deduce the observationally favored model of $\gamma$. 
\begin{itemize}
\item {\bf Chi-square $\boldsymbol {\chi^{2}}~$}:~Consider a set of observed data $\{ \mathcal{O}_{i}\}$ with possible errors $\{ \sigma_{i} \}$. The corresponding model estimates of the observed quantity is denoted by $\Omega_{i}(\gamma)$, where $\gamma$ is related to the scalar charge associated with each of the quasars. The $\chi^{2}$ of the distribution is then given by,
\begin{align}
\chi^{2} (\gamma) = \sum_{i} \frac{\{ \mathcal{O}_{i} - \Omega_{i}(\gamma) \}^{2}}{\sigma_{i}^{2}} \label{18}
\end{align}

\begin{table}[h]
%\tiny
\hfill{}

%{\centerline{\large Table 1}}
%{\centerline{Spin parameters of quasars corresponding to $r_{2}=0.1$ and $r_{2}=0$ (for comparison with GR) }}
\caption{The mass, accretion rate, optical and bolometric luminosity of the eleven quasars considered are reported. These are taken from \cite{Davis:2010uq}.}
\label{Table2}
%{\centerline{}}
%\vspace{-0.36cm}
\begin{center}
\centering
\begin{tabular}{|c|c|c|c|c|}

\hline
$\rm Object$ & $\mathcal{M}_0$ & $\rm log\dot{M}_0$ & $\rm log ~L_{opt}$ & $\rm log ~L_{bol}$ \\
\hline 
$\rm 0003+158$ & $\rm 9.16$ & $\rm 0.79$ & $\rm 45.87$ & $\rm 46.92 \pm 0.25$ \\ \hline
$\rm 1048-090$ & $\rm 9.01$ & $\rm 0.30$ & $\rm 45.45$ & $\rm 46.57 \pm 0.32$ \\ \hline
$\rm 1100+772$ & $\rm 9.13$ & $\rm 0.29$ & $\rm 45.51$ & $\rm 46.61 \pm 0.25$ \\ \hline
$\rm 1103-006$ & $\rm 9.08$ & $\rm 0.21$ & $\rm 45.43$ & $\rm 46.19 \pm 0.10$ \\ \hline
$\rm 1216+069$ & $\rm 9.06$ & $\rm 0.51$ & $\rm 45.62$ & $\rm 46.61 \pm 0.28$ \\ \hline
$\rm 1226+023$ & $\rm 9.01$ & $\rm 1.18$ & $\rm 46.03$ & $\rm 47.09 \pm 0.24$ \\ \hline
$\rm 1425+267$ & $\rm 9.53$ & $\rm 0.07$ & $\rm 45.55$ & $\rm 46.35 \pm 0.20$ \\ \hline
$\rm 1512+370$ & $\rm 9.20$ & $\rm 0.20$ & $\rm 45.48$ & $\rm 47.11 \pm 0.50$ \\ \hline
$\rm 1545+210$ & $\rm 9.10$ & $\rm 0.01$ & $\rm 45.29$ & $\rm 46.14 \pm 0.13$ \\ \hline
$\rm 1704+608$ & $\rm 9.29$ & $\rm 0.38$ & $\rm 45.65$ & $\rm 46.67 \pm 0.21$ \\ \hline
$\rm 2308+098$ & $\rm 9.43$ & $\rm 0.22$ & $\rm 45.62$ & $\rm 46.61 \pm 0.22$ \\ \hline
\end{tabular}
\hfill{}
\end{center}
\end{table}

In \ref{18}, $\sigma_{i}$ refers to the error associated with the observed optical luminosity for each of the quasars. It turns out that the error in optical luminosity is negligible compared to the error in bolometric luminosity which receives maximum contribution from the far-UV extrapolation as the uncertainty in the UV luminosity far exceeds other sources of error (e.g., optical or X-ray variability) \cite{Davis:2010uq}. Since the errors in optical luminosity of the quasars are not explicitly reported we consider the errors in the bolometric luminosity (reported in \cite{Davis:2010uq} and \ref{Table2}) as the maximum error possible in the estimation of the optical luminosity.

\begin{figure}[t!]
\centering
\includegraphics[scale=0.5]{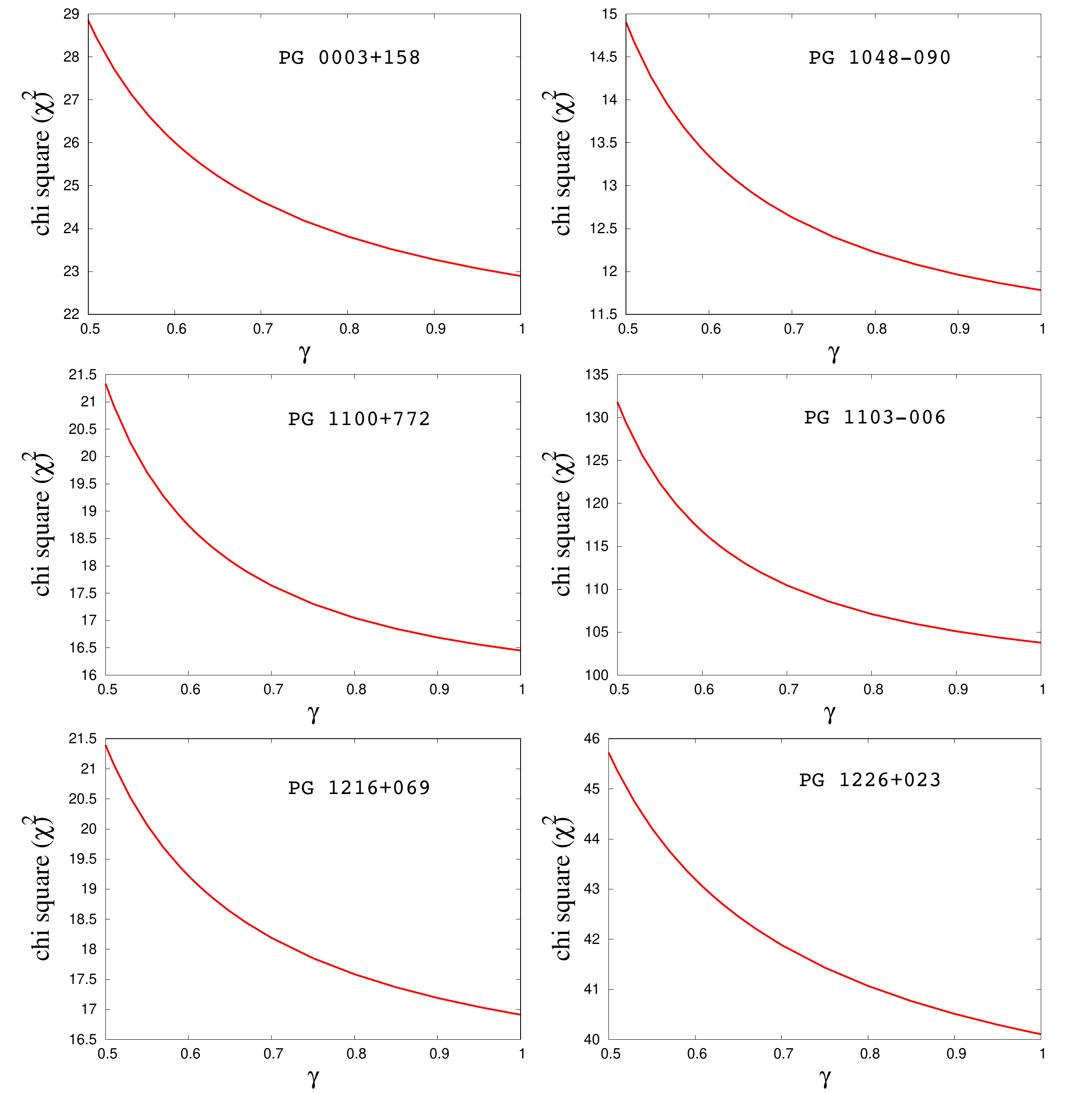}
\caption{ The above figure depicts variation of $\chi^{2}$ as a function of the metric parameter $\gamma$ for individual quasars with $M\geq 10^9 M_\odot$. For every quasar an uncertainty of $0.4 dex$ is considered in the mass while accretion rates are varied between $80\%$ to $140\%$ of the value reported in \cite{Davis:2010uq}, to compute the theoretical luminosity. It is evident from the plot that $\chi^{2}$ minimizes for $\gamma\sim 1$. For more discussion see text.}
\label{Fig_2a}
\end{figure}

\begin{figure}[t!]
\centering
\includegraphics[scale=0.5]{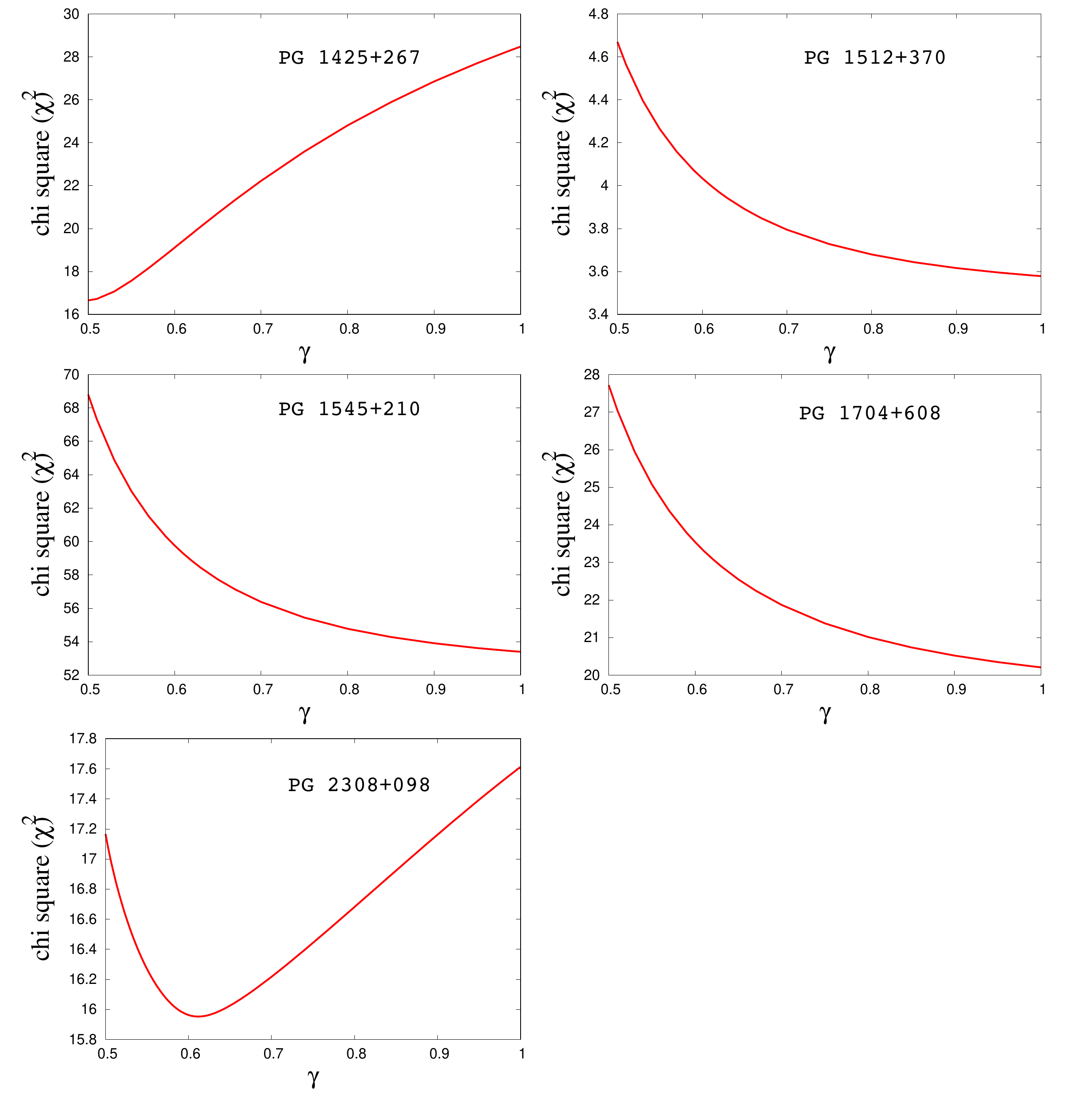}
\caption{The above figure depicts variation of $\chi^{2}$ as a function of the metric parameter $\gamma$ for individual quasars with $M\geq 10^9 M_\odot$. While computing the $\chi^{2}$ for a given value of $\gamma$ an uncertainty of $0.4 dex$ is considered in the mass estimates for all quasars while accretion rates are varied between $80\%$ to $140\%$ of the value reported in \cite{Davis:2010uq}, to compute the theoretical luminosity. The figure illustrates that for most quasars $\chi^{2}$ minimizes for $\gamma\sim 1$, the exceptions being PG 1425+267 and PG 2308+098.}
\label{Fig_2b}
\end{figure}

\begin{figure}[t!]
\centering
\includegraphics[scale=0.75]{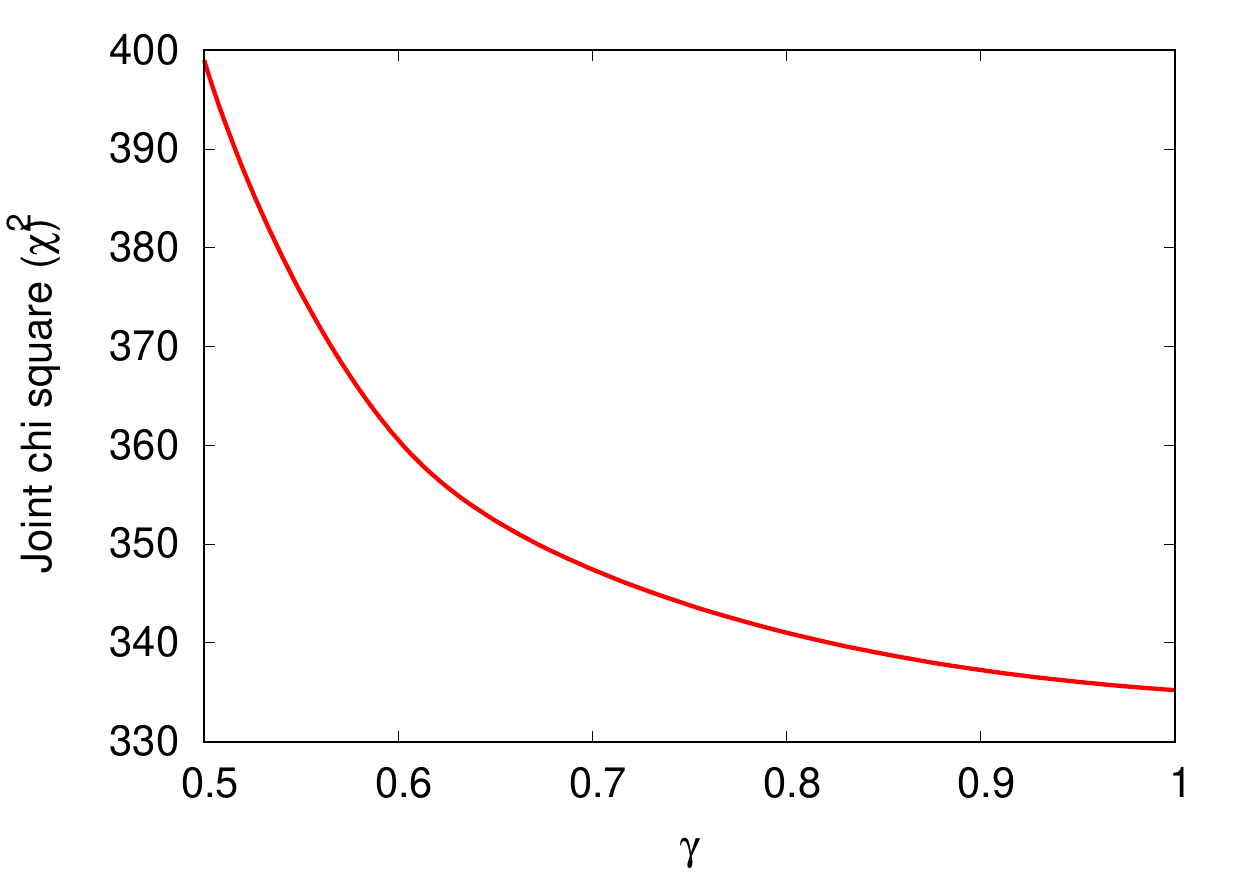}
\caption{The above figure depicts the joint $\chi^{2}$ (by summing the $\chi^{2}$ for all the quasars) as a function of the metric parameter $\gamma$. It is evident from the plot that $\chi^{2}$ minimizes for $\gamma\sim 1.0$.}
\label{Fig_2}
\end{figure}

In order to compute the the $\chi^{2}$, the theoretical estimate of optical luminosity $\Omega_{i}$ is required, which depends on the mass of the quasars, their accretion rates and the metric parameter $\gamma$ (which is related to the scalar charge associated with the black hole). As discussed in the last section we consider eleven quasars with mass $M\geq 10^9 M_\odot$ \cite{Davis:2010uq} in our analysis. For each of these quasars, the masses based on reverberation mapping are reported in \cite{Davis:2010uq} (also mentioned in \ref{Table2}) which are subject to systematic errors that override the statistical uncertainty in the input data. The systematic errors are difficult to quantify and a factor of $\sim 3 ~\rm (0.4 ~dex)$ error is considered as the characteristic uncertainty in the mass estimates \cite{Davis:2010uq}. For example, the mass of the quasar PG 1545 + 210 is taken to be $\textrm{log ~M}=9.10 \pm 0.4 M_\odot$. We denote the central value of the logarithm of the mass by $\mathcal{M}_0$ (reported in \ref{Table2}). Then in the logarithmic scale the mass $M$ of the quasars can vary between $\mathcal{M}_0-0.4 \lesssim \textrm{log ~M}\lesssim \mathcal{M}_0+0.4$. For example, for PG 1545 + 210, $\mathcal{M}_0=9.10$.\\
%For example, the mass of the quasar PG 1545 + 210 is taken to be $\textrm{log ~M}=9.10 \pm 0.4 M_\odot$. This systematic error of $0.4 ~dex$ is associated with the mass of all the eleven quasars \cite{Davis:2010uq}.\\
The accretion rates of these quasars can at most vary between $80\%$ to $140\%$ of the accretion rate reported in \ref{Table2} (see discussion in the last section), which can be used to compute the theoretical luminosity. 
For easier reference, the mass, accretion rate, optical and bolometric luminosities of the eleven PG quasars are reported in \ref{Table2}.

In order to compute the $\chi^2$ for a given source with central value of the logarithm of its mass $\mathcal{M}_0$, we first fix a value of $\gamma$ in the range 0.5 to 1. Then we fix the value of mass in the range $\mathcal{M}_0-0.4 \lesssim \textrm{log ~M}\lesssim \mathcal{M}_0+0.4$ and allow the accretion rate to vary between 0.8 to 1.4 times the accretion rate $\dot{M}_0$ reported in \ref{Table2} in steps of 0.1. For each different accretion rate, but fixed $\textrm{log ~M}$ and $\gamma$, we compute the theoretical optical luminosity and subsequently the $\chi^2$ as in \ref{18} and simply sum them up. While computing the $\chi^2$ the error in the observed optical luminosity $\sigma_i$ is required. As mentioned earlier in this section, $\sigma_i$ is taken to be the error in the bolometric luminosity (reported in \ref{Table2}) as the maximum error possible in the estimation of the optical luminosity. \\
This method therefore considers the effect of variation in the accretion rate.

Next we consider a different value of $\textrm{log ~M}$ for the same quasar in the allowed range mentioned above, keep the $\gamma$ fixed, but vary the accretion rate as before, compute the resultant $\chi^2$ and again add them up to the previous sum of $\chi^2$. We repeat this procedure for all values of $\textrm{log ~M}$ in the aforesaid range, where the stepsize of varying $\textrm{log ~M}$ is also taken to be 0.1.
In this way $\chi^2$ for a particular magnitude of $\gamma$ is calculated which is essentially the sum of the $\chi^2$ obtained by varying the mass and the accretion rate.

Now the last two steps are repeated for all $\gamma \in \left(0.5,1\right)$ to obtain the variation of $\chi^2$ with $\gamma$ for the given source. Subsequently, the above process is reiterated for all the eleven quasars which gives the dependence of the $\chi^{2}$ on $\gamma$ for the individual quasars. In \ref{Fig_2a} and \ref{Fig_2b} we plot the variation of $\chi^{2}$ with $\gamma$ for each of the eleven quasars. We note that for most of the quasars the $\chi^{2}$ minimizes for $\gamma\approx 1$ except for PG 1425+267 and PG 2308+098. This indicates that the Schwarzschild scenario is mostly favored by optical observations of quasars compared to the Janis Newman Winicour spacetime.

We next compute the joint chi-square by summing the $\chi^{2}$ of all the quasars for a given value of $\gamma$, and repeating this process for all $\gamma$ in the physically allowed range $\gamma \in \left(0.5,1\right)$. This is depicted in \ref{Fig_2} which illustrates that the total $\chi^{2}$ minimizes for $\gamma\approx 1$, thereby favoring the Schwarzschild scenario. In order to strengthen our conclusions we consider a few more error estimators. In the remaining error estimators, the theoretical luminosity is computed with masses of the quasars from \ref{Table2} while accretion rates considered are $1.4$ times the accretion rate reported in \ref{Table2}. This is because we are considering the quasars in the high mass end ($M\geq 10^9 M_\odot$) \cite{Davis:2010uq}
(discussion in \ref{S3-1}).

\begin{figure}[htp]
\subfloat[Nash Sutcliffe Efficiency \label{Fig_3a}]{\includegraphics[scale=0.65]{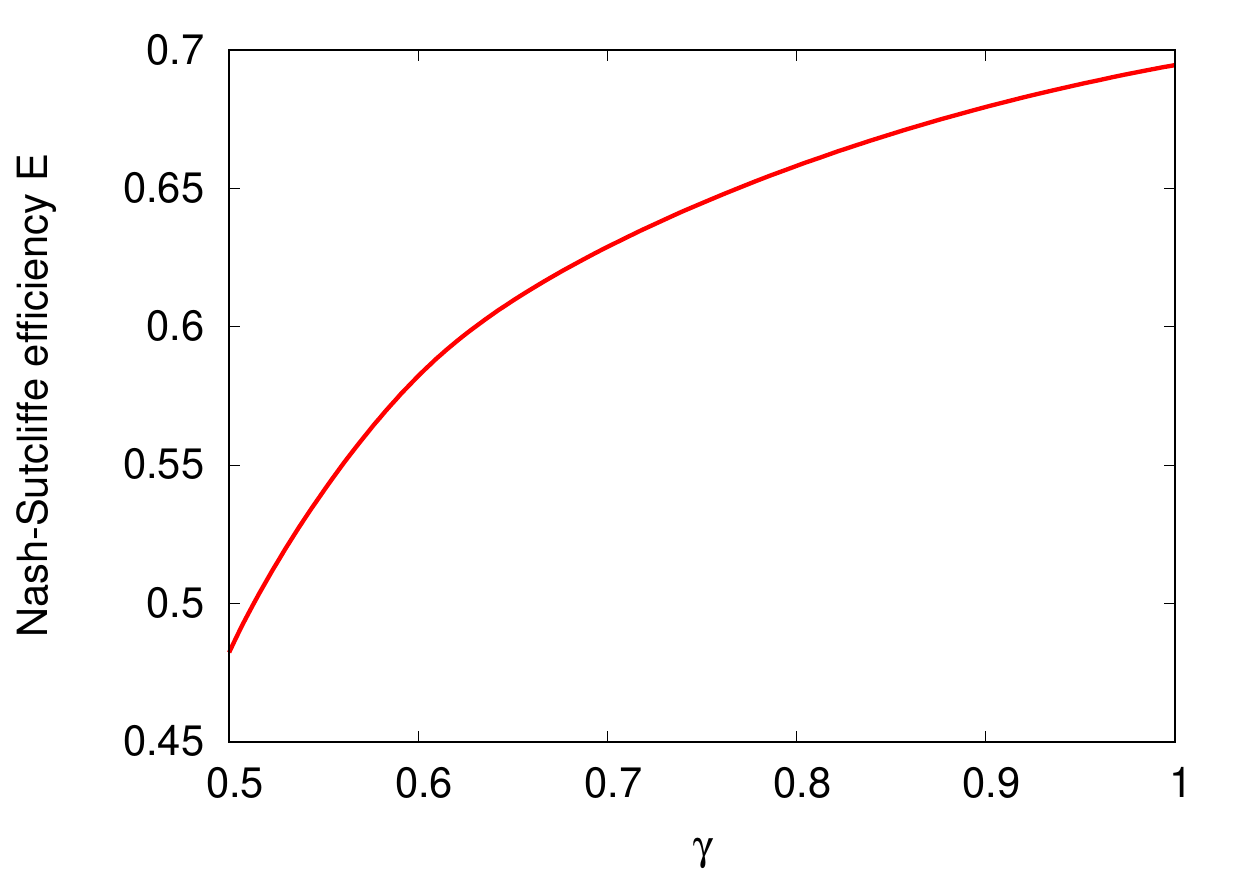}}
\subfloat[Modified form of Nash Sutcliffe Efficiency\label{Fig_3b}]{\includegraphics[scale=0.65]{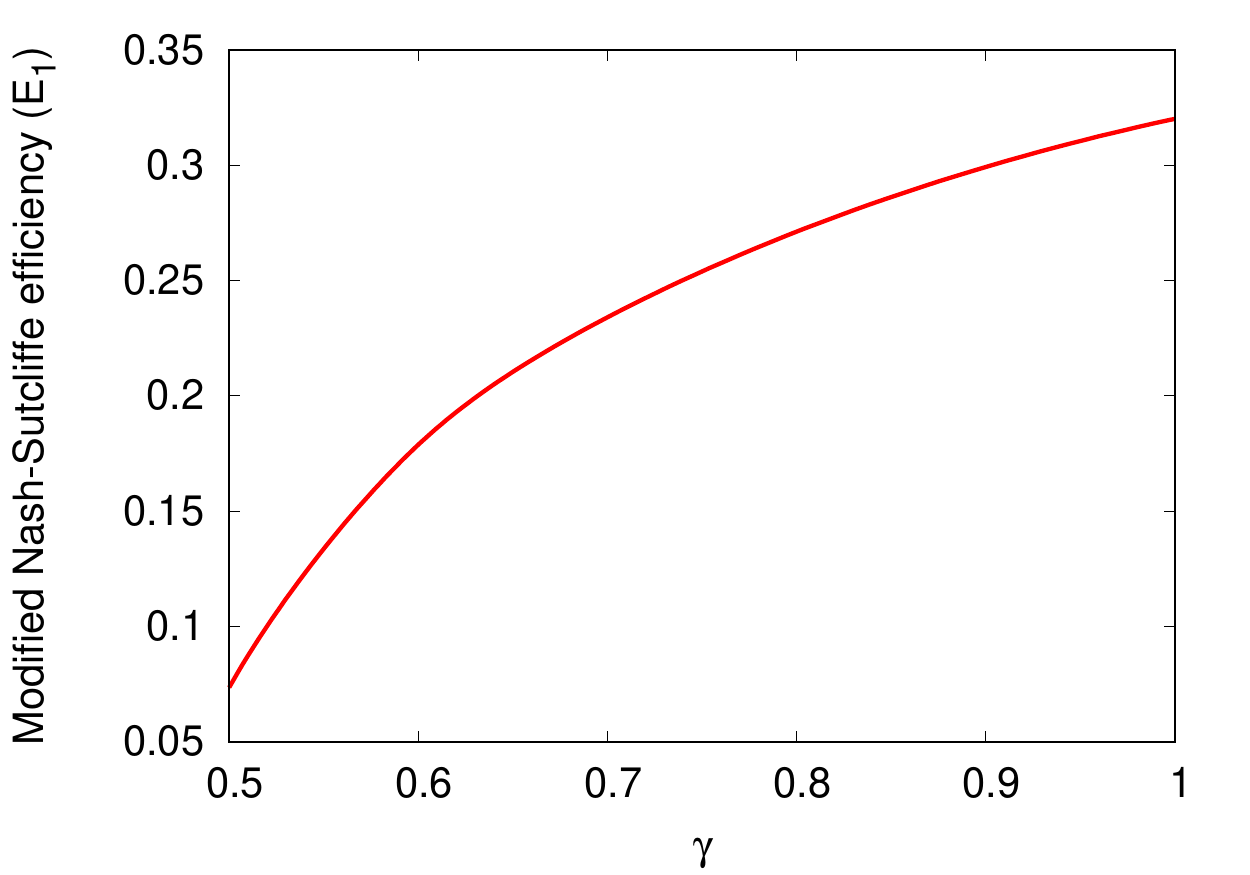}}
\caption{The above figure depicts variation of (a) the Nash-Sutcliffe Efficiency $E$ and (b) the modified form of the Nash-Sutcliffe Efficiency $E_1$ with the metric parameter $\gamma$. Both the error estimators maximize for $\gamma\sim 1$.}
%\label{Fig_3}
\end{figure}

\item \textbf{Nash-Sutcliffe Efficiency and its modified form:} Nash-Sutcliffe Efficiency $E$ \cite{NASH1970282,WRCR:WRCR8013,2005AdG.....5...89K} is related to the sum of the squared differences between the observed and the predicted values normalized by the variance of the observed values.  This error estimator assumes the form, 
\begin{align}
E(\gamma) = 1 - \frac{\sum_{i} \{ \mathcal{O}_{i} - \Omega_{i}(\gamma) \}^{2}}{\sum_{i} \{ \mathcal{O}_{i} - \mathcal{O}_{av} \}^2} \label{19}
\end{align}
where $\mathcal{O}_{\rm av}$ denotes average of the observed values of the optical luminosities of the quasars. Unlike $\chi^{2}$, the model which best describes the observation maximises the Nash-Sutcliffe Efficiency. A model with $E\sim 1$ is considered to be an ideal model that accurately predicts the observations. While calculating the theoretical optical luminosity for a given $\gamma$ in \ref{19}, the masses of the quasars are considered from Table 1 of \cite{Davis:2010uq} while the accretion rates are multiplied by a factor of $1.4$ since we are considering the quasars in the high mass end ($M\geq 10^9 M_\odot$) \cite{Davis:2010uq}. This choice of mass and accretion rate is taken for every quasar in the remaining error estimators we discuss further.

As depicted in \ref{Fig_3a} in our case, $E$ maximizes for $\gamma \sim 1$, indicating that the Schwarzschild scenario predicts the observation better than the Janis Newman Winicour background.
\par
Nash-Sutcliffe Efficiency $E$ tends to be oversensitive to higher values of the luminosity for taking square of the error in the numerator (see e.g. \ref{19}). 
Therefore, a modified version of the Nash-Sutcliffe Efficiency denoted by $E_1$ \cite{WRCR:WRCR8013} is used, where,
\begin{align}
\label{20}
E_{1}(\gamma)&=1-\frac{\sum_{i}|\mathcal{O}_{i}-\Omega_{i}(\gamma)|}{\sum _{i}|\mathcal{O}_{i}-\mathcal{O}_{\rm av}|}
\end{align}
Similar to $E$, a model which maximizes $E_1$ is considered to be a better description of the data. \ref{Fig_3b} illustrates that $E_1$ maximizes for $\gamma\sim 1$. The conclusions drawn from these two error estimators corroborate our previous findings.

\begin{figure}[htp]
\subfloat[Index of agreement \label{Fig_4a}]{\includegraphics[scale=0.65]{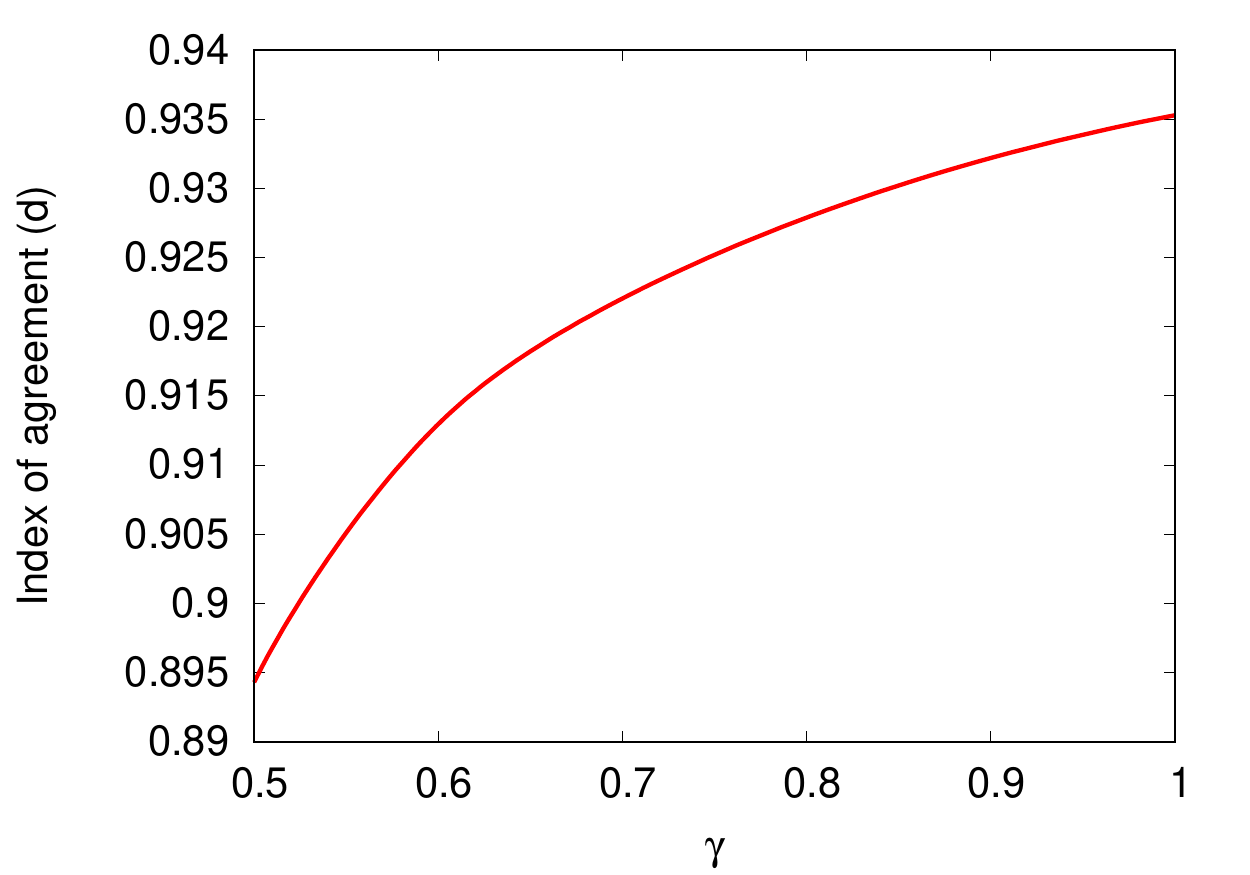}}
\subfloat[Modified index of agreement\label{Fig_4b}]{\includegraphics[scale=0.65]{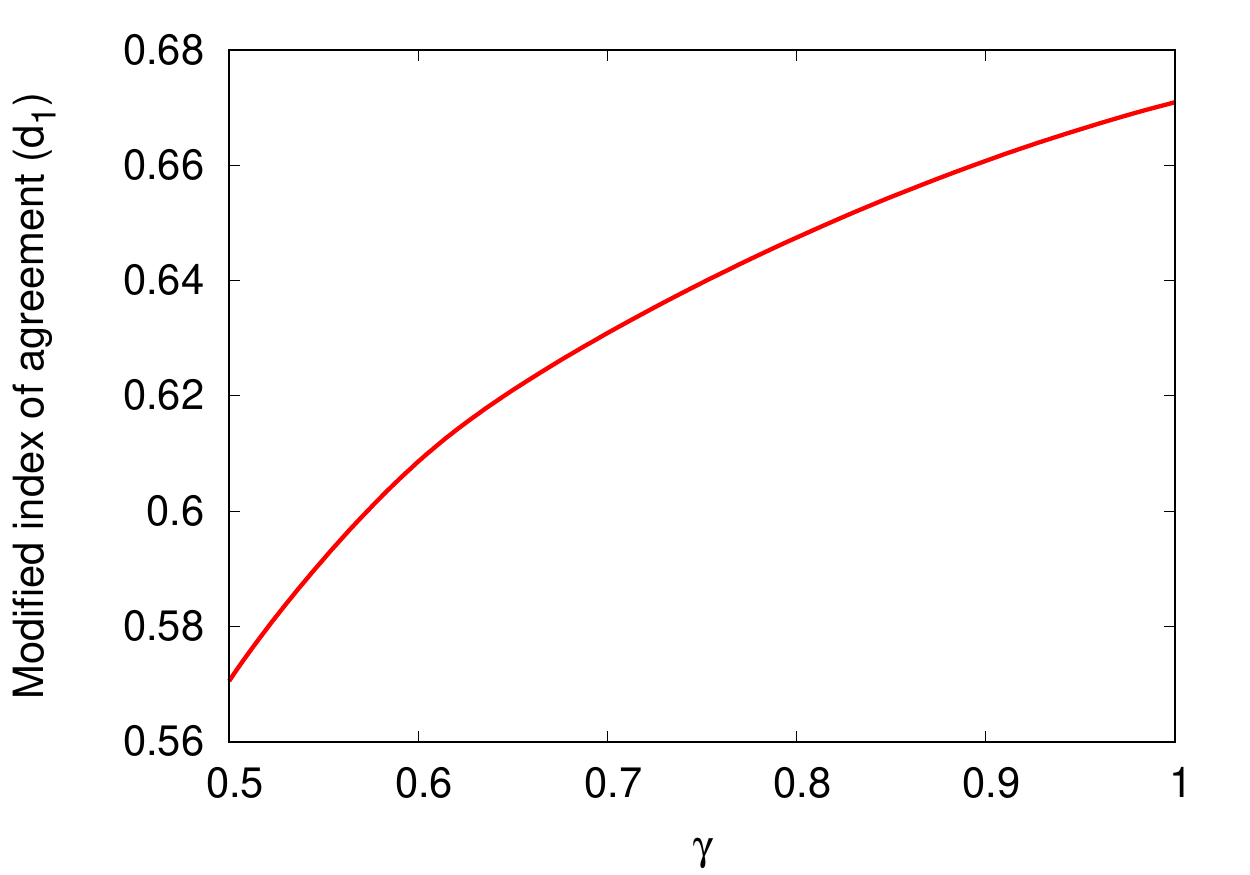}}
\caption{The above figure depicts variation of (a) index of agreement $d$ and (b) the modified index of agreement ${\bf d_1}$ with the metric parameter $\gamma$. Both the error estimators maximize for $\gamma\sim 1$ favoring the Schwarzschild scenario.}
\label{Fig_4}
\end{figure}

\item \textbf{Index of agreement and its modified form:} The index of agreement was proposed \cite{willmott1984evaluation, doi:10.1080/02723646.1981.10642213,2005AdG.....5...89K} to overcome the insensitivity of the Nash-Sutcliffe Efficiency and its modified form towards the differences between the observed and predicted means and variances. Denoted by $d$, it assumes the form,
\begin{align}
d(\gamma) = 1 - \frac{\sum_{i} \{ \mathcal{O}_{i} - \Omega_{i}(\gamma) \}^{2}}{\sum_{i} \{ |\mathcal{O}_{i} - \mathcal{O}_{av}| + |\Omega_{i}(\gamma) - \mathcal{O}_{av}| \}^{2}} \label{21}
\end{align} 
The denominator, which  denotes the maximum deviation of each pair of observed and predicted luminosities from the average luminosity is known as the potential error.

Similar to Nash-Sutcliffe Efficiency, the index of agreement suffers from oversensitivity to higher values of optical luminosity due to the presence of the squared luminosities in the numerator of \ref{21} and hence its  modified version $d_1$ is proposed, where,
\begin{align}
d_{1}(\gamma) = 1 - \frac{\sum_{i}|\mathcal{O}_{i} - \Omega_{i}(\gamma)|}{\sum_{i} \{ |\mathcal{O}_{i} - \mathcal{O}_{av}| + |\Omega_{i}(\gamma) - \mathcal{O}_{av}| \}} \label{22}
\end{align}
Similar to the previous error estimators, we note from \ref{Fig_4a} and \ref{Fig_4b} that the model which best describes the observation maximizes $d$ and $d_1$ and hence corresponds to $\gamma\sim 1$. Therefore, the conclusions drawn previously remain unaltered, i.e. the Schwarzschild scenario seems to be favored by optical observations of quasars compared to the Janis Newman Winicour spacetime.

\end{itemize}

\section{Conclusion}
\label{S5}
The main goal of this work is to explore the characteristics of electromagnetic observations in the Janis-Newman-Winicour spacetime and confront them with the available observational data. This naturally involves investigating the nature of the black hole shadow and accretion in this background. Below we enlist the important results of this work:
\begin{itemize}
\item While investigating the properties of the shadow, we note that the presence of the scalar charge decreases the effects of the gravitational lensing and diminishes the shadow radius compared to the Schwarzschild scenario. With the increase in scalar charge or decrease in $\gamma$ the radius of the photon sphere increases while that of the shadow decreases which is one of the unique properties of the Janis-Newman-Winicour spacetime. 
A spinning black hole also casts a smaller shadow compared to a Schwarzschild black hole, although the scalar charge causes a greater reduction in the shadow radius compared to the Kerr scenario. This feature can partially remove the degeneracy between the JNW metric parameter $\gamma$ and the spin, from the recently observed image of M87*. Given the uncertainty in the mass estimates of the object, the observed angular diameter of M87* can be reproduced within the error bars, both by $\gamma\geq 0.57$ and any magnitude of the Kerr parameter (\ref{f6}). However, the Schwarzschild scenario explains the observed shadow for most of the allowed values of $M$ and in this way the Schwarzschild scenario is more favored by the observed shadow of M87*. Also extreme values of $\gamma$ ($\gamma <0.57$) are completely ruled out by the first image of the black hole, purely based on its small angular diameter. Moreover,  M87* also exhibits a powerful jet with $P_{jet}\geq 10^{42} \rm erg ~s^{-1}$ and $|a| \geq 0.5$ is required to explain the requisite jet power \cite{Akiyama:2019fyp}. This estimate of $|a|$ is also consistent with the shadow related observation, given the uncertainties in its mass estimate. Therefore, the observed jet further corroborates \gr\ over the JNW spacetime.  
A future observation of a black hole viewed at a high inclination angle and having precise and independent estimations of its mass and distance can be further used to establish/falsify the viability of the JNW spacetime.

\item In the Winicour solution when the metric parameters $\gamma$ and $b$ are treated as independent, a new regime emerges where $b$ is negative and $\gamma \le -0.5$. This represents a horizonless compact object with real positive solutions for photon sphere and shadow. This is an interesting generalization in the parameter space of the Janis-Newman-Winicour spacetime which has not been discussed much in the literature.

\item Apart from studying the nature of the shadow in the Janis-Newman-Winicour spacetime, we also explore the effects of this background on the accretion onto the compact object. We compute the theoretical estimates of optical luminosity from the accretion disk for a sample of Palomar Green quasars with $M\geq 10^9 M_\odot$ \cite{Davis:2010uq} and compare them with the corresponding observations. The uncertainties associated with the mass and the accretion rates are taken into account while computing the theoretical luminosity from the accretion disk which are subsequently used to evaluate the error estimators. For every allowed magnitude of mass and accretion rate, the variation of $\chi^2$ with $\gamma$ is computed for all the quasars. It turns out that the $\chi^2$ minimizes in the Schwarzschild scenario for most of the quasars, thereby favoring \gr\ over the JNW background. This is eventually followed by evaluating the joint-$\chi^2$ and other error estimators like the Nash-Sutcliffe efficiency, the index of agreement etc., which in turn supports our earlier findings. This result is also in agreement with the first observed shadow of a black hole which is another independent window to test the nature of strong gravity in the electromagnetic domain.

It is however important to mention that the quasars are multicomponent systems containing the accretion disk, the corona, the jet and the dusty torus emitting in all bands of the electromagnetic spectrum and we have not explicitly fitted the observed SED with the Novikov-Thorne model which mimicks the emission only from the accretion disk. We are interested in disentangling the effect of the background metric from the SED and only emissions from regions very close to the black hole gets modified by the background spacetime. This is one of the primary reason we choose to model the accretion disk since the effect of the metric on the other components is not so important. Secondly, modelling the entire spectral energy distribution (SED) theoretically is extremely challenging since it depends not only on the background spacetime but also on the properties of the accretion flow and one often resorts to phenomenological models to address this issue. Discerning the effect of the metric from the SED therefore becomes quite non-trivial.
Our goal in this work is not to model the entire SED but to constrain the value of $\gamma$ from the accretion observations using a theoretical model for the disk. Amongst the available theoretical models the Novikov-Thorne model is very successful in explaining the emission from the accretion disk and our work is simply a first attempt to identify the observationally favored value of the scalar charge of the JNW spacetime from the accretion data.

\end{itemize}

\section*{Acknowledgements}
The research of SSG is partially supported by the Science and Engineering
Research Board-Extra Mural Research Grant No. (EMR/2017/001372), Government of India.
The research of S.S is funded by CSIR, Government of India.

%%%%%%%%%%%%%%%%%%%%%%%%%%%
\bibliography{P3,accretion,KN-ED,Black_Hole_Shadow}
\bibliographystyle{utphys1}
%\bibliography{x.bib}
%\bibliographystyle{plain}
%\bibliographystyle{JHEP}
%\bibliographystyle{plainnat}
\end{document}